%% file: main.tex
\pdfoutput=1
\documentclass[acmsmall,screen,nonacm]{acmart}

\input{preamble}

\begin{document}

\title{Underpinning Property-Based Testing}

\author{Ernest Ng}
\orcid{0009-0007-3706-6336}
\affiliation{%
  \institution{Amazon Web Services}
  \city{Boston}
  \state{MA}
  \country{USA}
}
\affiliation{%
  \institution{Cornell University}
  \city{Ithaca}
  \state{NY}
  \country{USA}
}
\email{eyn5@cornell.edu}

\author{Segev Elazar Mittelman}
\orcid{0009-0000-0516-4694}
\affiliation{%
  \institution{University of Maryland}
  \city{College Park}
  \state{MD}
  \country{USA}
}
\affiliation{%
  \institution{Amazon Web Services}
  \city{Arlington}
  \state{VA}
  \country{USA}
}
\email{segevem@umd.edu}

\author{Joshua M. Cohen}
\orcid{0000-0002-9555-8781}
\affiliation{%
  \institution{Amazon Web Services}
  \city{Boston}
  \state{MA}
  \country{USA}
}
\email{cohenjo@amazon.com}

\author{Katherine Hough}
\orcid{0000-0003-4397-1635}
\affiliation{%
  \institution{Amazon Web Services}
  \city{Arlington}
  \state{VA}
  \country{USA}
}
\email{kmhough@amazon.com}

\author{Michael Hicks}
\orcid{0000-0002-2759-9223}
\affiliation{%
  \institution{Amazon Web Services}
  \city{Arlington}
  \state{VA}
  \country{USA}
}
\affiliation{%
  \institution{University of Pennsylvania}
  \city{Philadelphia}
  \state{PA}
  \country{USA}
}
\email{mwh@cis.upenn.edu}

\author{Harrison Goldstein}
\orcid{0000-0001-9631-1169}
\affiliation{%
  \institution{University at Buffalo, SUNY}
  \city{Buffalo}
  \state{NY}
  \country{USA}
}
\email{hgoldste@buffalo.edu}

\begin{abstract}
As the property-based testing community strives to automate challenging parts of the process away from users~---~in particular, the construction of generators~---~it is critical that we can clearly state what it means for that automation to be correct.
We present a new framework for PBT, implemented in the Lean theorem prover, that provides a more solid foundation to build on.
Our generators are interpretable: interpreted one way, they can be run efficiently as programs that sample randomness; and interpreted another way, they enable a wide range of correctness proofs, including proofs about \emph{distributions} that other foundational PBT models do not support.
We evaluate our framework extensively, with both small and large benchmarks and case studies, arguing that the extra power brought by distributional proofs is practically important for the next generation of generator automation.
\end{abstract}

\maketitle

\section{Introduction}\label{sec:intro}

\epigraph{\textbf{underpin} \emph{trans.} To support or strengthen\ldots\ especially by laying a solid foundation\ldots\ or by substituting stronger or more solid for weaker or softer materials.}{\textit{Oxford English Dictionary}}

\noindent
Property-based testing (PBT) combines software testing, which is critical for ensuring that software systems behave as expected, with formal methods, which give a robust language for formally stating those expectations.
Rather than checking a function's output on one particular input, like standard unit testing, PBT checks that a \emph{property} holds over a whole class of inputs.
When testing compilers, module APIs, and databases, for example, that class might be well-typed programs, well-formed method-call sequences, or valid SQL queries.
Valid inputs like these are often extremely rare in the space of all inputs, so testing such properties requires \emph{constrained random generation}: efficiently sampling from the inputs in that class.

To address the challenge of constrained random generation, the PBT literature uses \emph{generators}: programs specifically designed to efficiently sample from an appropriate distribution of test data.
Well-designed generators have been used to uncover interesting bugs in complex targets like compilers~\cite{palkaTestingOptimisingCompiler2011,disselkoenHowWeBuilt2024}, stateful services~\cite{jaber2026high}, and high-assurance automotive software~\cite{artsTestingAUTOSARSoftware2015},
and can outperform weaker generators enhanced with feedback-guided fuzzing~\cite{lampropoulosCoverageGuidedProperty2019,li2025havoc,vikramFuzzingAgentsGenerators2026}.

Needing to hand-write good generators is a usability barrier for PBT~\cite{goldsteinPropertyBasedTestingPractice2024}, so recent work has begun to automate the generator creation process: systems like Cobb~\cite{lafontaineWeveGotYou2025}, Palamedes~\cite{goldsteinSearchConstrainedRandom2026}, and QuickChick~\cite{elazarmittelmanTestingTheoremsFully2026,paraskevopoulouComputingCorrectlyInductive2022} automate the production of generators for a wide range of use cases.
AI programming tools can extend that range further~\cite{vikramFuzzingAgentsGenerators2026}.
Unfortunately, existing approaches for generator synthesis offer, at best, partial guarantees that the generators they produce are \emph{high quality}.
Existing tools can show that a generator is \emph{sound} and \emph{complete}: it is able to generate all and only the values that satisfy the chosen set of constraints.
More than that, though, a generator's \emph{distribution} must be useful for finding bugs.
For example, if a generator could theoretically produce every natural number, but in practice it produces \leaninline{0} half of the time and has a vanishingly small chance of producing values greater than \leaninline{30}, then it will miss bugs that manifest only with larger numbers.
Modern frameworks for synthesizing generators and proving their correctness say nothing about distributions; indeed, Palamedes will produce a skewed \li{Nat} generator like the one just described.
It is possible to address distributional problems through experimentation and tuning, e.g., using a tool like Tyche~\cite{goldsteinTycheMakingSense2024}, but not always~---~if a user wants to rule out some very rare bad behavior, for example, they may need to test their generator for as long as or longer than the system under test.
Ultimately, since developers use tests to detect defects in their code, poorly distributed generators constitute a real risk to software correctness.

In this paper, we address these challenges with a new library called \ourlib,
written in Lean~\cite{demouraLean4Theorem2021}, that can express distributional correctness properties to ensure that both hand-written and synthesized PBT generators are high-quality.
\ourlib provides a foundational semantics, just like the original foundation for PBT, QuickChick~\cite{paraskevopoulouFoundationalPropertyBasedTesting2015}, but uses a richer \emph{probabilistic} semantics, rather than QuickChick's \emph{possibilistic} one.
This means that clients can prove distributional properties about their generators, e.g., that a generator produces the uniform distribution, that the expected size of a generated value is less than a certain bound, or that the probability of a certain feature exceeds a minimum threshold.
\ourlib generators are also more expressive than prior encodings: they can express potentially non-terminating generators, which arise naturally (even a simple generator for arbitrary natural numbers is not structurally terminating), and users can prove almost-sure termination as needed.

\ourlib is not just a formal framework; it is a fully featured library designed for practical use.
Thanks to a tagless-final embedding~\cite{caretteFinallyTaglessPartially2009}, the same code that is used for proofs is directly executable.
Leveraging recent work on boosting PBT generator speed~\cite{richeyFailFasterStaging2026}, these generators run at least as fast as flagship PBT frameworks like Haskell's QuickCheck and OCaml's Base\_quickcheck (\autoref{sec:performance}).%
\footnote{\ourlib clients need not be written in Lean: \ourlib generators can use Lean's FFI to interface with code in other languages, or even directly invoke external services via \li{IO} API calls.}
We show that automatic generator synthesis tools, specifically Palamedes and Specimen~\cite{StrataorgSpecimen2026} (a port of QuickChick to Lean), can be re-tooled to construct \ourlib generators, and we use this new underpinning to automatically derive correctness proofs for Specimen generators.
Finally, we implement generators for two real-world systems in \ourlib.
We use Specimen to derive \ourlib generators for the Cedar policy language~\cite{cutlerCedarNewLanguage2024}, proving distributional properties desired by the Cedar development team, and we use LLMs to implement generators for the Strata intermediate verification language~\cite{StrataorgStrata2026}, proving correctness and distributional properties about them, thereby demonstrating that our proof infrastructure is legible to AI-based proof tools.
Ultimately, we position \ourlib as a new underpinning for PBT, with the theoretical strength and practical engineering necessary to build trustworthy test infrastructure.

In the rest of this paper, we present the following contributions:
\begin{itemize}
  \item The design and implementation of \ourlib, a generator library with a probabilistic semantics which supports efficient and potentially non-terminating generators (\autoref{sec:basalt} and \autoref{sec:recursion}).
  \item Proof infrastructure for reasoning about the support, distributions, termination, and cost of generators (\autoref{sec:proofs}).
  \item An evaluation (\autoref{sec:case-studies}) that benchmarks \ourlib's performance, shows that \ourlib can underpin synthesis tools Palamedes and Specimen, and applies \ourlib to two real-world systems, Cedar and Strata.
\end{itemize}
We conclude with related (\autoref{sec:related}) and future (\autoref{sec:conclusion}) work.

\section{\ourlib Basics}\label{sec:basalt}

In this section, we describe the core of \ourlib, which allows users to write generators that support both execution and distributional proofs.

\subsection{\ourlib by Example}\label{sec:example}

Let us look at two \ourlib generators to get a sense of the pieces that they are built from.
The simplest possible generator looks like this:
\begin{lean}
def genFiveToTen [Generator G] : G Nat := choose 5 10
\end{lean}
This generator uniformly samples a natural number between 5 and 10 (inclusive).

As with QuickCheck~\cite{claessenQuickCheckLightweightTool2000}, \ourlib generators can be combined using monadic \leaninline{do}-notation:
\begin{lean}
def genPairFiveToTen [Generator G] : G (Nat × Nat) := do
  let x ← genFiveToTen
  let y ← genFiveToTen
  return (x, y)
\end{lean}
The back-arrow notation can be read as ``sample from the generator on the right and bind it to the variable on the left.''

\subsection{The Generator Type Class}\label{sec:type-class}

The example generator definitions may look a bit mysterious; \leaninline{genFiveToTen} produces a value of type \leaninline{G Nat}, but \leaninline{G} itself is not defined.
Indeed, the definitions are polymorphic over the choice of \leaninline{G}~---~\ourlib generators use a typed tagless-final~\cite{caretteFinallyTaglessPartially2009} embedding, where generators are defined only by the operations that they support.

Concretely, a generator is any type constructor that can implement the following simplified type class (see \autoref{sec:recursion} for the full version):
\begin{lean}
class Generator (g : Type → Type) extends
  Monad g,
  RandomChoice g,
  ...
\end{lean}
A generator must be a \li{Monad}, which enables the QuickCheck-style \li{do}-notation.
A generator also needs a way to make random choices: the \leaninline|RandomChoice| class provides it as a single operation, \leaninline|choose|, which makes an inclusive choice over a nonempty range of natural numbers.
\begin{lean}
class RandomChoice (m : Type u → Type v) where
  choose : (lo hi : Nat) → (h : lo ≤ hi) → m Nat
\end{lean}
Every other way of making random choices is built on top of \leaninline{choose}.

The benefit of a tagless-final embedding is that the same program can be interpreted in multiple ways.
This allows us to write one generator and use it twice: once for testing and once for proofs of correctness.
It works because the explicit type of a generator like \li{genPairFiveToTen} is:
\begin{lean}
genPairFiveToTen : ∀ {G : Type → Type} [Generator G], G (Nat × Nat)
\end{lean}
We can read this as saying ``if you give me a type constructor \li{G} and show that \li{G} has all of the operations of a \li{Generator}, then I can give you a generator of type \leaninline{G (Nat × Nat)}.''
In the next two subsections, we present \li{Generator} instances for testing and proof, respectively.

\subsection{Generator Instance: IO}\label{sec:executable}

To run a generator during testing, we use the \li{IO} instance of \li{Generator},
which does what you might expect: the monad instance sequences effectful computations, and the \li{RandomChoice} instance samples a PRNG.
We can use Lean's \li{#eval} command to sample from the generators in \autoref{sec:example} by instantiating \li{G} with \li{IO}:
\par\smallskip\noindent
\begin{minipage}[t]{0.45\textwidth}
\begin{lean}[aboveskip=0pt,belowskip=0pt]
#eval (genFiveToTen : IO Nat)
-- 7
\end{lean}
\end{minipage}\hfill
\begin{minipage}[t]{0.55\textwidth}
\begin{lean}[aboveskip=0pt,belowskip=0pt,xleftmargin=0pt]
#eval (genPairFiveToTen : IO (Nat × Nat))
-- (10, 6)
\end{lean}
\end{minipage}
\par\medskip

\begin{figure}[b]
  \centering
  \includegraphics[height=9\baselineskip]{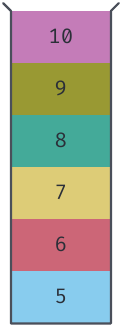}
  \caption{\li{genFiveToTen}, drawn as a jar of sand.}\label{fig:genFiveToTen}
  \Description{A full jar holding six layers of colored sand of equal height, labeled 5 through 10 from bottom to top.}
\end{figure}

\subsection{Generator Instance: SPMF}

To prove properties about a generator, including ones about its distribution, we use the \li{SPMF} instance of \li{Generator}.
This instance interprets generators as \emph{sub-probability mass functions} (SPMFs), implemented in Lean as follows:%
\footnote{Measure-theoretic representations would be necessary for uncountable domains, but the domains we care about for testing are countable so (sub-)probability mass functions suffice.}
\begin{lean}
def SPMF (α : Type) : Type := {μ : α → ℝ≥0∞ // (∑' a, μ a) ≤ 1}
\end{lean}
The mass function \leaninline{μ} assigns a mass (essentially a probability) to every value in the domain \leaninline{α}.%
\footnote{We use extended, non-negative reals to represent mass because they are used by related theories in Lean's Mathlib.}
And we require that the sum of the mass over all \leaninline{α}s be at most 1.

To build the interpretation, we can start with \li{RandomChoice}.
We express a random choice as a uniform distribution over a range:%
\footnote{In the real artifact, \li{choose} returns a subtype that guarantees \li{n} is in the appropriate range, so the \li{if} check is not needed.}
\begin{lean}
instance : RandomChoice SPMF where
  choose lo hi h :=
    let k := hi - lo + 1
    ⟨fun n => if (lo ≤ n ∧ n ≤ hi) then (1 / k) else 0, ...⟩
\end{lean}
The angle-bracket notation pairs a value, in this case our mass function, with a proof, in this case the proof that the total mass is bounded above by 1. We omit the proof for brevity.
We discuss our reasoning for this particular randomness primitive in \autoref{app:primitives-for-randomness}.

Then we can define the monad operations as the standard discrete probability monad~\cite{ramsey_pfeffer_2002, erwig_kollmansberger_2005}:
\begin{lean}
instance : Monad SPMF where
  pure a := ⟨fun a' => if a' = a then 1 else 0, ...⟩
  bind x f := ⟨fun b => ∑' a, x a * f a b, ...⟩
\end{lean}
The \li{pure} operation produces a point-mass distribution, and \li{bind} sequences two distributions, averaging the continuation \li{f} over outcomes of \li{x}, weighted by their probabilities.

In this paper we visualize SPMFs as jars of colored sand, where each color corresponds to a different outcome.
Since the total mass of all outcomes is at most one, the jar cannot be filled past the top.
To sample a value, imagine picking a random spot in the jar and taking the outcome whose color is at that spot.
\autoref{fig:genFiveToTen} visualizes \li{genFiveToTen}.

\subsection{Faithfulness}\label{sec:faithfulness}

A key benefit of \ourlib is that both the IO and SPMF interpretations run the same code, but that does not necessarily mean that they must perfectly agree.
One might hope we can make this argument by parametricity, but in Lean it is technically possible for a generator to inspect types.
To get around this, we prove \emph{faithfulness}~---~agreement between the code that runs and the code for proofs~---~on a per-generator basis.
We describe this proof process at a high level in \autoref{app:proving-faithfulness}, and the full machinery appears in the artifact.

\section{Recursive Generators in \ourlib}\label{sec:recursion}

In this section, we complete the \li{Generator} type class, and discuss how \ourlib represents recursive, potentially non-terminating generators.

\subsection{More Examples}

\begin{figure}[t]
\hfill
  \begin{minipage}[t]{0.43\textwidth}
\begin{lean}[xleftmargin=0pt]
def genNat [Generator G] : G Nat :=
  oneOf [
    fun () => return 0,
    fun () => do
      let n ← genNat
      return (n + 1)
  ]
partial_fixpoint
\end{lean}
  \end{minipage}%
\hfill
  \begin{minipage}[t]{0.47\textwidth}
\begin{lean}[xleftmargin=0pt]
def genHeap [Generator G]
    (lo : Nat) : G (Tree Nat) :=
  oneOf [
    fun () => return leaf,
    fun () => do
      let x := lo + (← genNat)
      return (node (← genHeap x) 
        x (← genHeap x))
  ]
partial_fixpoint
\end{lean}
  \end{minipage}%
  \hfill\null
  \caption{Recursive generators: \li{genNat} produces arbitrary natural numbers; \li{genHeap} produces valid min-heaps.}\label{fig:genNat-genHeap}
  \Description{Two Lean code listings side by side. The left listing defines genNat and the right listing defines genHeap.}
\end{figure}

\autoref{fig:genNat-genHeap} shows two recursive generators.
The first, \li{genNat}, produces an arbitrary natural number.
At each step, the generator chooses from two options with equal probability (the \li{oneOf} combinator uses \li{choose} to choose uniformly between branches).
It may return \li{0}, or else it may generate a natural number \li{n} by calling \li{genNat} recursively and then add \li{1} to it.

The second, \li{genHeap}, produces valid min-heaps (trees whose values never decrease along any root-to-leaf path)~\cite{cormenIntroductionAlgorithms2022}.
Again, the generator chooses from two options with equal probability.
It may return a \leaninline{leaf}, ending generation.
Or it may compute a new value \leaninline{x}, by adding a random delta to \leaninline{lo} (the minimum allowable value in the sub-heap), and recursively generate two sub-heaps whose new minimum value is \leaninline{x}.
The result is a generator that produces all possible valid heaps, and only valid heaps.

\subsection{Relaxing Structural Termination}

Neither \li{genNat} nor \li{genHeap} terminates structurally, i.e., by recursing on strictly smaller inputs.
Instead, these generators terminate \emph{almost surely} (i.e., with probability 1): \li{genNat} stops at each step with probability $\frac{1}{2}$, and for every pair of children that \li{genHeap} creates, one of them will end up a \li{leaf} in expectation.
This feature of \ourlib makes it possible to avoid programming around and reasoning about \emph{fuel}.
In QuickChick, \li{genHeap} would instead need to take a \li{fuel} argument and return an \li{Option} (see \autoref{app:fuel}).
The \li{fuel} argument ensures that the generator always has some decreasing argument for its recursive calls.
This \li{fuel} management causes a number of paper cuts: the user needs to read and write code that is unrelated to the generator's actual operation; the result is wrapped in an \li{Option}, so downstream users need to handle out-of-fuel failures; end-users need to decide on an amount of fuel to pass (which could potentially be insufficient); and proofs need to reason about termination and correctness all at once.

Most importantly, indexing generators with fuel makes distributional arguments much more complicated.
For example, consider a generator that does rejection sampling to filter \li{genPairFiveToTen} down to a uniform distribution over those pairs that are in sorted order:
\begin{lean}
def genPairFiveToTenSorted [Generator G] : G (Nat × Nat) := do
  let (x, y) ← genPairFiveToTen
  if x ≤ y then return (x, y) else genPairFiveToTenSorted
partial_fixpoint
\end{lean}
This generator is clearly not structurally terminating, but it does terminate ``almost surely,'' since each attempt succeeds with a fixed, positive probability.
The generator also obviously gives the distribution we want, with an equal chance of producing any sorted pair.
If we added fuel, the distribution would no longer be strictly uniform; it would only be uniform when conditioned on returning \li{some}.

\subsection{The Full Generator Type Class}

To support generators that do not terminate structurally, the \li{Generator} class needs a few more pieces:
\begin{lean}
class Generator (g : Type → Type) extends
  Monad g,
  RandomChoice g,
  (∀ α, Inhabited (g α)),
  (∀ α, CCPO (g α)),
  MonoBind g
\end{lean}
Lean requires \li{Inhabited}, \li{CCPO}, and \li{MonoBind} in order to use its \li{partial_fixpoint} feature.%
\footnote{This feature is explained in Section 7.6 of the Lean Language Reference~\cite{LeanReference435-rc2} and in a couple of blog posts by Joachim Breitner~\cite{ExtrinsicTerminationProofs,F91LeanBlog}.}

Lean's standard library defines a \li{CCPO} as follows:
\begin{lean}
def chain (c : α → Prop) : Prop := ∀ x y , c x → c y → x ⊑ y ∨ y ⊑ x
class CCPO (α : Sort u) extends PartialOrder α where
  has_csup {c : α → Prop} (hc : chain c) : ∃ s, ∀ x, s ⊑ x ↔ (∀ y, c y → y ⊑ x)
\end{lean}
Given a partial order (\leaninline{⊑}), a \li{chain} is a totally ordered subset of values under that partial order.
If all chains in the partial order have least upper bounds, then we say that partial order is ``chain-complete''~---~i.e., it is a CCPO.

To understand this paper's contributions, we can focus on a relatively narrow application of this idea: CCPOs give a way to build fixpoints of functions.
Every CCPO has a least element \leaninline{⊥}, the least upper bound of the empty chain.
If a function \li{f} is \emph{monotone} (i.e., \leaninline{∀ x y, x ⊑ y → f x ⊑ f y}), then \li{f} has a least fixed point.
To see why, imagine applying \li{f} to \leaninline{⊥} over and over; since \leaninline{⊥ ⊑ f ⊥} and \li{f} is monotone, this gives a chain:
\begin{center}
  \leaninline{⊥ ⊑ f ⊥ ⊑ f (f ⊥) ⊑ f (f (f ⊥)) ⊑ ...}
\end{center}
Each element of the chain approximates the fixed point more closely, and in many cases the least upper bound of the chain (which is guaranteed to exist) is the fixed point itself.
The \li{partial_fixpoint} annotation (which appears in \li{genNat} and \li{genHeap}) uses \li{Lean.Order.fix} to represent non-terminating recursion; we show what these fixed points look like for SPMFs below.

The \li{Inhabited} constraint is required by \li{partial_fixpoint}, and the \li{MonoBind} constraint asserts that the \li{Monad}'s bind operation is \emph{monotone} with respect to the partial order.

\subsection{Revisiting IO}

The \li{IO} instances for \li{CCPO}, \li{Inhabited}, and \li{MonoBind} are all provided by Lean in the standard library.
Lean compiles \li{partial_fixpoint} definitions to ordinary \li{partial} functions that do not interact with Lean's kernel.
We can sample these generators just like their non-recursive counterparts:

\par\smallskip\noindent
\begin{minipage}[t]{0.38\textwidth}
\begin{lean}[aboveskip=0pt,belowskip=0pt]
#eval (genNat : IO Nat)
-- 4
\end{lean}
\end{minipage}\hfill
\begin{minipage}[t]{0.62\textwidth}
\begin{lean}[aboveskip=0pt,belowskip=0pt,xleftmargin=0pt]
#eval (genHeap 0 : IO (Tree Nat))
-- node (node leaf 4 leaf) 2 (node leaf 3 leaf)
\end{lean}
\end{minipage}
\par\medskip

\subsection{Revisiting SPMFs}

The \li{CCPO} instance for SPMFs is much more interesting.
Recall that the total mass of an SPMF is allowed to be less than one.
If the probabilities of all terminating outcomes sum to less than one, then we can interpret the remaining mass as the probability of \emph{no terminating outcome}~---~in other words, divergence.

We can construct a \li{CCPO} instance that builds on this intuition.
First, we need a partial order that orders distributions by how well defined they are.
Concretely:
\begin{lean}
instance : Lean.Order.PartialOrder (SPMF α) where
  rel p q := ∀ a, p a ≤ q a
  ...
\end{lean}
In other words, \li{p} is less than \li{q} when \li{q} assigns at least as much mass to every \li{a} as \li{p} does; the probability of an event can only go up along a chain.

\begin{wrapfigure}[10]{r}{0.25\textwidth}
  \centering
  \captionsetup{justification=centering}
  \includegraphics[height=7\baselineskip]{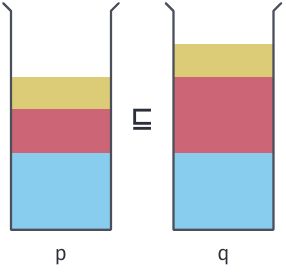}
  \caption{Two jars with \li{p} $\sqsubseteq$ \li{q}.}\label{fig:sand-order}
  \Description{Two partly filled jars of sand labeled p and q. Both have the same three colors, but q has more of the middle color.}
\end{wrapfigure}
To build some intuition, recall the jars of sand from \autoref{fig:genFiveToTen}.
So far, every jar has been full, but now we consider what happens when there is some empty space at the top; an unfilled jar represents an SPMF whose total mass is less than one.
If we pick a spot in the empty space, we get no value at all; in other words, the generator diverges.
If jars \li{p} and \li{q} are ordered by the relation above, then we can get from \li{p} to \li{q} by only adding sand, never removing any.
A chain is then a sequence of jars, each with at least as much sand of each color as the one before it.
If we want to know what the least upper bound of the chain looks like, we fill a jar with the maximum of each color of sand that appears in any jar in the chain.

Put back into Lean, the \li{CCPO} instance for \li{SPMF} looks like:
\begin{lean}
instance : CCPO (SPMF α) where
  has_csup := fun c => ⟨⟨fun a => ⨆ μ ∈ c, μ a, ...⟩, ...⟩
\end{lean}
In other words, the least upper bound of a chain is the \li{SPMF} where the mass of each outcome is the supremum of its masses along the chain.

\begin{figure}[H]
  \centering
  \includegraphics[height=9\baselineskip]{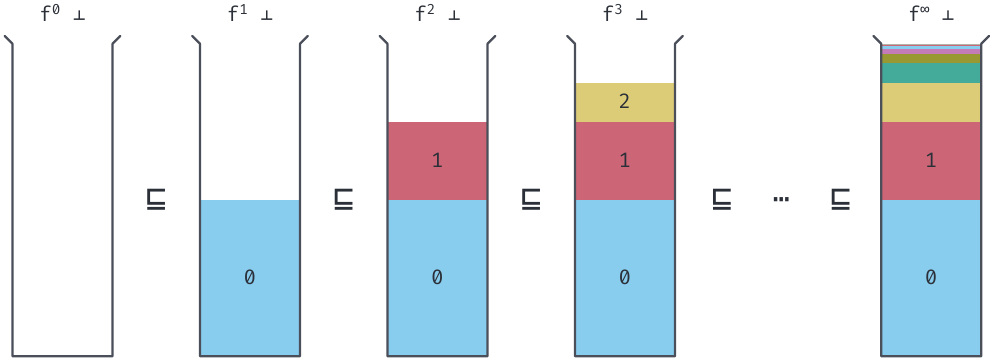}
  \caption{The chain of more-defined distributions formed by iterating \li{f} from the definition of \li{genNat}.}\label{fig:genNat-chain}
  \Description{A sequence of rectangular diagrams, each representing a sub-probability mass function, arranged in a chain separated by a less-than operator.}
\end{figure}

Let us revisit \li{genNat}, now written out explicitly with \li{Lean.Order.fix}:
\begin{lean}
def genNatExplicit [Generator G] : G Nat :=
  let f := fun g => oneOf [
      fun () => return 0,
      fun () => do let n ← g; return (n + 1)
    ]
  Lean.Order.fix f
\end{lean}
This generator either stops with probability $\frac{1}{2}$ and produces \li{0} or else it recursively generates a number \li{n} and returns \li{n + 1}.
Interpreted at \li{SPMF}, the iteration chain of \li{f} starts from the empty jar \leaninline{⊥}, as shown in \autoref{fig:genNat-chain}: each application of \li{f} adds sand for one more outcome, never removing any.
The least upper bound of this chain assigns probability $\frac{1}{2^{n+1}}$ to each \li{n}, and it is a fixed point of \li{f}: it is exactly \li{genNat}'s distribution.
Practically, this means that we can work with this definition in Lean safely, even though it does not terminate structurally.

\subsection{Size and Backtracking}\label{sec:extensions}

\ourlib also supports two common patterns from real PBT generators: automatic handling of \emph{sizes} and infrastructure for \emph{backtracking}.
Rather than support these in \li{Generator}, we layer them on top and give each its own proof infrastructure: \ourlib provides sizes through an auxiliary \li{Sized} type class and backtracking via the \li{OptionT} monad transformer.
See \autoref{app:size} and \autoref{app:backtracking} for more detail.

\section{Proving Properties of Generators}\label{sec:proofs}

So far we have seen the underlying representation of \ourlib generators, along with a few ways to interpret them.
Now, we lean more heavily on the \li{SPMF} interpretation specifically to enable distribution-aware proofs about generators.
We discuss a variety of properties that developers can prove (\autoref{sec:support-proofs} and \autoref{sec:distribution-proofs}), and our approach to proof automation (\autoref{sec:program-logics}).
Finally, we discuss the trade-offs between proofs and sampling for generator validation (\autoref{sec:alternatives-to-proof}).

\subsection{Proofs about Support}\label{sec:support-proofs}

The first two properties that we care about proving for a generator are well known in the literature: soundness and completeness~\cite{paraskevopoulouFoundationalPropertyBasedTesting2015,dybjerCombiningTestingProving2003,dybjerRandomGeneratorsDependent2004,vanderrestCompletelyUniqueAccount2022}.
To define these properties, we first need to define support in the context of SPMFs; the support of a generator is the set of values with nonzero mass:
\begin{lean}
def support (p : SPMF α) : Set α := {x | p x ≠ 0}
\end{lean}
Then a generator is \emph{sound} if every \li{a} in the support of \li{g} satisfies some predicate \li{P}.
\begin{lean}
def IsSound (g : SPMF α) (P : α → Prop) : Prop :=
  ∀ a, a ∈ SPMF.support g → P a
\end{lean}
If we think of \li{P} as the \emph{precondition} of a property under test, then soundness says that \li{g} will not violate the property's precondition and cause false positives.
And a generator is \emph{complete} if every \li{a} that satisfies \li{P} is in the support of \li{g}.
\begin{lean}
def IsComplete (g : SPMF α) (P : α → Prop) : Prop :=
  ∀ a, P a →  a ∈ SPMF.support g
\end{lean}
If \li{g} cannot produce some values that satisfy \li{P}, then no matter how many tests we run we may fail to exercise some aspects of the target system.
We often combine these properties into one, \li{IsSoundAndComplete}.
We can prove that \li{genHeap} is sound and complete with respect to \li{isHeap}, a definition of the min-heap invariant.
\begin{lean}
theorem genHeap.sound_complete : 
  IsSoundAndComplete (genHeap lo) (isHeap lo)
\end{lean}
For this and all other theorems in this section, we give the full proof scripts in \autoref{app:proofs}; most of these proofs are quite short, thanks to automation that we discuss in \autoref{sec:program-logics}.

\subsection{Proofs about Distributions}\label{sec:distribution-proofs}

Soundness and completeness are expressible in other PBT frameworks; the properties in this section require a semantic domain like SPMFs to express and prove.

\paragraph{Probabilities}

The quintessential property that becomes provable with our SPMF semantics is the probability of a particular outcome.
\begin{lean}
def ProbabilityOf (g : SPMF α) (a : α) : ℝ≥0∞ := g a
\end{lean}
This allows us to characterize parts of the distribution, potentially highlighting issues that could impact testing.
We can prove that \li{genHeap} puts half of its mass on the empty heap:
\begin{lean}
theorem genHeap.prob_empty : ProbabilityOf (genHeap lo) leaf = 1 / 2
\end{lean}
This is quite bad!
The empty heap is not useful for testing, and thus in practice we might want to \emph{tune} \li{genHeap} to ensure that it produces a more interesting distribution.

\paragraph{Expected Values}

We can also talk about expected values over the distribution of a generator.
Concretely, if we are given a function \li{f} over the values that \li{g} generates, we can compute an expected value as follows:
\begin{lean}
def ExpectedValue (g : SPMF α) (f : α → ℝ≥0∞) : ℝ≥0∞ := ∑' a, g a * f a
\end{lean}
Choosing \li{f} lets us measure various aspects of the distribution, for example, the average size of a generated value or the average number of times a certain constructor appears.
We can prove that the expected value of \li{genNat} is at most 1:
\begin{lean}
theorem genNat.expect_le : ExpectedValue genNat (fun n => n) ≤ 1
\end{lean}
This tells us that \li{genNat} mostly produces small numbers; in fact, it is exactly the skewed generator described in \autoref{sec:intro}.
Like \li{genHeap}, it might need some tuning to be actually usable for testing.

\paragraph{Almost-Sure Termination}

Since we allow generators that do not terminate structurally, we might also want to prove \emph{almost-sure termination}.
We define \li{mass} as the sum of the masses of the values \li{g} can generate, and then we say a generator almost surely (AS) terminates if that mass is 1:
\begin{lean}
def mass (p : SPMF α) : ℝ≥0∞ := ∑' a, p a
def IsAlmostSurelyTerminating (g : SPMF α) : Prop := SPMF.mass g = 1
\end{lean}

AS termination might seem net-neutral, since it fixes a problem that we created for \ourlib.
But remember that allowing generators that do not terminate structurally brings them more in line with the way QuickCheck generators are naturally written, enables more flexible distributional control, and avoids the overhead of fuel in termination-insensitive proofs.

\paragraph{Cost}
Another correctness criterion that \ourlib unlocks is efficiency.
Two generators that have the same support can achieve that support in very different ways.
For example, \li{genHeap} produces valid heaps directly, ``by construction,'' but one could also generate heaps by generating arbitrary trees until a heap happens to show up.
We provide the following predicate for expressing efficiency in terms of the relationship between generated values and the number of choices the generator makes to produce them:
\begin{lean}
def IsCostBounded (g : SPMF.Cost α) (c : α → Nat) : Prop :=
  ∀ (a, n) ∈ SPMF.support g, n ≤ c a
\end{lean}
This requires a new interpretation of generators, \li{SPMF.Cost}, where \li{SPMF.Cost α} is equivalent to \li{SPMF (α × Nat)}.
The definition is the same as \li{SPMF}, but every \li{choose} adds 1 to the running total (the \li{Nat}), and \li{bind} adds up the totals of the computations it sequences.
We can show that our \li{genHeap} makes $O(n + m)$ choices, where $n$ is the size of the tree and $m$ is the sum of the values in the tree:
\begin{lean}
theorem genHeap.cost_bounded :
    IsCostBounded (genHeap lo) (fun t => 3 * t.size + t.sum + 1)
\end{lean}
The cost model keeps us honest: looking only at the structure of \li{genHeap}, we might conclude that it makes a number of choices linear in the size of the tree, but each call to \li{genNat} also makes a number of choices linear in the value it produces, so we have to account for those costs as well.

\paragraph{Positive AS Termination}

We can combine many of the above properties~---~AS termination, expectation, and cost~---~to prove a stronger property about generator efficiency.
AS termination says that a generator eventually terminates, but this does not guarantee that it does so in a reasonable amount of time.
We can also prove \emph{positive} AS termination, which additionally requires the expected running time to be finite.
A generator is positively AS terminating if it is AS terminating and its expected cost is bounded by some finite \li{B}:
\begin{lean}
def IsExpectedCostBounded (g : SPMF.Cost α) (B : ℝ≥0∞) : Prop :=
  ExpectedValue g (fun (_, n) => n) ≤ B
\end{lean}
Producing \li{n} with \li{genNat} takes $n + 1$ choices, and we just saw that the expected value of \li{genNat} is at most 1.
So we expect \li{genNat} to make at most 2 choices on average, and indeed we can prove it:
\begin{lean}
theorem genNat.expected_cost_bounded :
    IsExpectedCostBounded genNat 2
\end{lean}
Together with \li{genNat.terminates} (\autoref{app:proofs}), this shows that \li{genNat} is positively AS terminating.

Not every AS terminating generator is positively AS terminating.
At each step, \li{genHeap} either stops with probability $\frac{1}{2}$ or recurses twice.
On average, each call to \li{genHeap} makes exactly one more recursive call.
We can prove that no finite bound holds:
\begin{lean}
theorem genHeap.not_expected_cost_bounded {B : ℝ≥0∞} (hB : B ≠ ⊤) (lo : Nat) :
    ¬ IsExpectedCostBounded (genHeap lo) B
\end{lean}
Combined with \li{genHeap.prob_empty}, this paints a particularly poor picture of \li{genHeap}: half of its heaps are empty, and yet generating a heap takes unboundedly many choices on average.
In the language of branching processes~\cite{harris1963theory}, \li{genHeap} is \emph{critical}: it terminates with probability 1, but its expected cost is infinite.

\subsection{Compositional Proofs via Program Logics}\label{sec:program-logics}

The previous subsections outline \emph{what} users might prove with \ourlib, but not \emph{how} they prove those things.
Some of the proofs are straightforward and can be done easily with Lean's built-in proof automation, but others are tedious without extra help.

We define a tactic called \li{walk} (which appears in many of the proofs in \autoref{app:proofs}) that implements \emph{verification condition generation} based on weakest preconditions~\cite{dijkstraGuardedCommandsNondeterminacy1975} for a wide range of theorems about generators.
For each shape of theorem, we implement a lightweight program logic: soundness uses a demonic nondeterminism semantics~\cite{morgan_et_al_1996,zilbersteinDemonicOutcomeLogic2025}; completeness uses an angelic semantics that follows a style similar to incorrectness logic~\cite{ohearnIncorrectnessLogic2019} and coverage types~\cite{zhouCoveringAllBases2023}; termination uses an expectation logic that is real-valued rather than truth-valued~\cite{mciverAbstractionRefinementProof2004,kozenProbabilisticPDL1985}; and cost bounding works like soundness at a different interpretation.
We give an example of this tactic in action in \autoref{app:walk-example}.

We chose to build our own machinery here, rather than rely on Lean's built-in verification condition generation tactic~\cite{LeanReference435-rc2} or one of the emerging Lean program verifiers like Loom~\cite{gladshteinFoundationalMultiModalProgram2026} and VCVio~\cite{cryptoeprint:2026/899}, to give ourselves maximal control.
Lean's built-in machinery only works over \li{Prop}-valued assertions, and all of the existing machinery handles termination differently from how we do.
As these systems mature, we will revisit our approach to automation.

We also understand that many users will prefer to write correctness proofs like the ones in this section with the help of AI.
In \autoref{sec:case-studies}, we describe some real-world workflows that use both our \li{walk} tactic and AI assistance to build generator correctness proofs; we see no need for users to have to choose one or the other.

\subsection{Proofs and Testing: Better Together}\label{sec:alternatives-to-proof}

If a developer is not ready to prove properties of their generator yet, they can also \emph{test} their generators to get a sense of how they perform.
Existing work has explored this approach in detail, and \ourlib provides commands for getting statistical feedback from generators that is similar to what libraries like QuickCheck and Hypothesis~\cite{maciverHypothesisNewApproach2019} provide.
A user could also pipe their generator statistics into a tool like Tyche~\cite{goldsteinTycheMakingSense2024} to make the statistical feedback visual.

Users of \ourlib may develop hypotheses about distributional properties using testing and then choose to prove those properties to get stronger guarantees.
\ourlib can also prove properties outside the reach of testing, answering questions such as:
\begin{itemize}
  \item \emph{Can this generator, in principle, generate every value I care about?}
  Issues of completeness cannot be validated by sampling, by definition.
  This kind of issue appears in the literature: \citet{LeoThesis2018} discusses how when testing information-flow control designs, a generator for indistinguishable machine states missed a large part of the state space.
  \item \emph{Are there rare cases where this generator produces a bad (invalid, extremely large, etc.) value?}
  If an event is rare, the user may need to test their generator for longer than their system under test to be confident that bad events will not come up in practice.
  \item \emph{Does a generator with parameters behave a certain way for all possible parameters?}
  If a generator is parameterized (e.g., by a size bound, by data, or even by another generator), it is impossible to sample it without first picking values for its parameters; this means it is difficult to establish certain properties in a vacuum.
  \citet{palkaTestingOptimisingCompiler2011}'s generator for well-typed lambda terms, for example, is parameterized by a generator for types.
  Sampling can only validate properties about the two generators together, while proofs can decouple those properties.
  \item \emph{Is the distribution of this generator equal to one I care about?}
  For applications where specific distributions are critical (e.g., when a precise distribution of values is needed to evaluate performance), sampling cannot give a firm guarantee of distributional equivalence.
  Some tools in the literature target uniform distributions~\cite{claessenGeneratingConstrainedRandom2015, canou_darrasse_2009}, while others attempt to match arbitrary user-defined distributions~\cite{tjoaTuningRandomGenerators2025,zhuangAriadneAutomaticallyTuning2026}; none of these tools provide proofs.
  \item \emph{Does the generator terminate and is the expected size of generated values bounded?}
  Sampling a generator that fails to terminate, or that produces arbitrarily large values, can slow down development and lead to frustrating feedback loops.
  As discussed above, generators like \li{genHeap} have infinite expected cost and may rarely produce extremely large values; this shows up in the literature as well~\cite{canou_darrasse_2009,mistaBranchingProcessesQuickCheck2018}.
\end{itemize}
We recommend a balanced approach, combining testing for quick feedback with proofs for generators as they evolve over time.

\section{Scaling up \ourlib}
\label{sec:case-studies}

So far, we have seen that \ourlib enables new kinds of validation for PBT generators; in this section, we show how it can scale to real-world workflows.
First, we argue that \ourlib is a competitive choice based on performance: we show that \ourlib generators run at least as fast as generators in established PBT frameworks (\autoref{sec:performance}).
Second, we show that \ourlib can serve as a foundational substrate for generator automation by retargeting two existing generator-synthesis libraries, Palamedes~\cite{goldsteinSearchConstrainedRandom2026} and Specimen~\cite{StrataorgSpecimen2026} (a Lean port, and expansion, of QuickChick~\cite{lampropoulosGeneratingGoodGenerators2017}), to use \ourlib (\autoref{sec:ports}).
This closes the loop discussed in \autoref{sec:intro}: users can synthesize a generator, tune it, and guarantee that it is high-quality.
Finally, we present two industrial case studies of \ourlib.
We used it to develop generators for well-typed programs in Cedar, an open-source authorization language used widely in industry (\autoref{sec:cedar}).
We also used \ourlib to develop generators for well-typed programs in Strata, an intermediate verification language that is under active development at a major technology company (\autoref{sec:strata}).
Development of the proofs and some of the generators in these case studies was greatly accelerated by AI-assisted programming tools, demonstrating a compelling vision for how \ourlib can scale in practice.

\subsection{Run-Time Performance}
\label{sec:performance}

\ourlib is designed not just to prove properties about generators, but also to execute them on real systems.
For generator execution to be practical, it must be fast. (Indeed, a recent PBT user study reveals that industry PBT practitioners run tests frequently with a limited time budget~\cite{goldsteinPropertyBasedTestingPractice2024}.)
Inspired by recent work on improving PBT generator performance~\cite{richeyFailFasterStaging2026}, we implemented two optimizations that ensure \ourlib generators run efficiently.
First, rather than use Lean's standard \li{IO.rand}, which we found introduced several unnecessary allocations for every random draw, we implemented the SplitMix64 PRNG~\cite{steeleFastSplittablePseudorandom2014} in C and linked it into a standalone Lean library that \ourlib calls.%
\footnote{\citet{richeyFailFasterStaging2026} found that using the SplitMix PRNG as the source of randomness for PBT generators yielded significant generator performance improvements.}
Second, we implemented common generator combinators like \li{oneOf} and \li{frequency} as macros that expand to \li{if}-chains; this avoids allocating lists of generators in a tight loop.

\begin{figure}[ht]
\includegraphics[width=\textwidth]{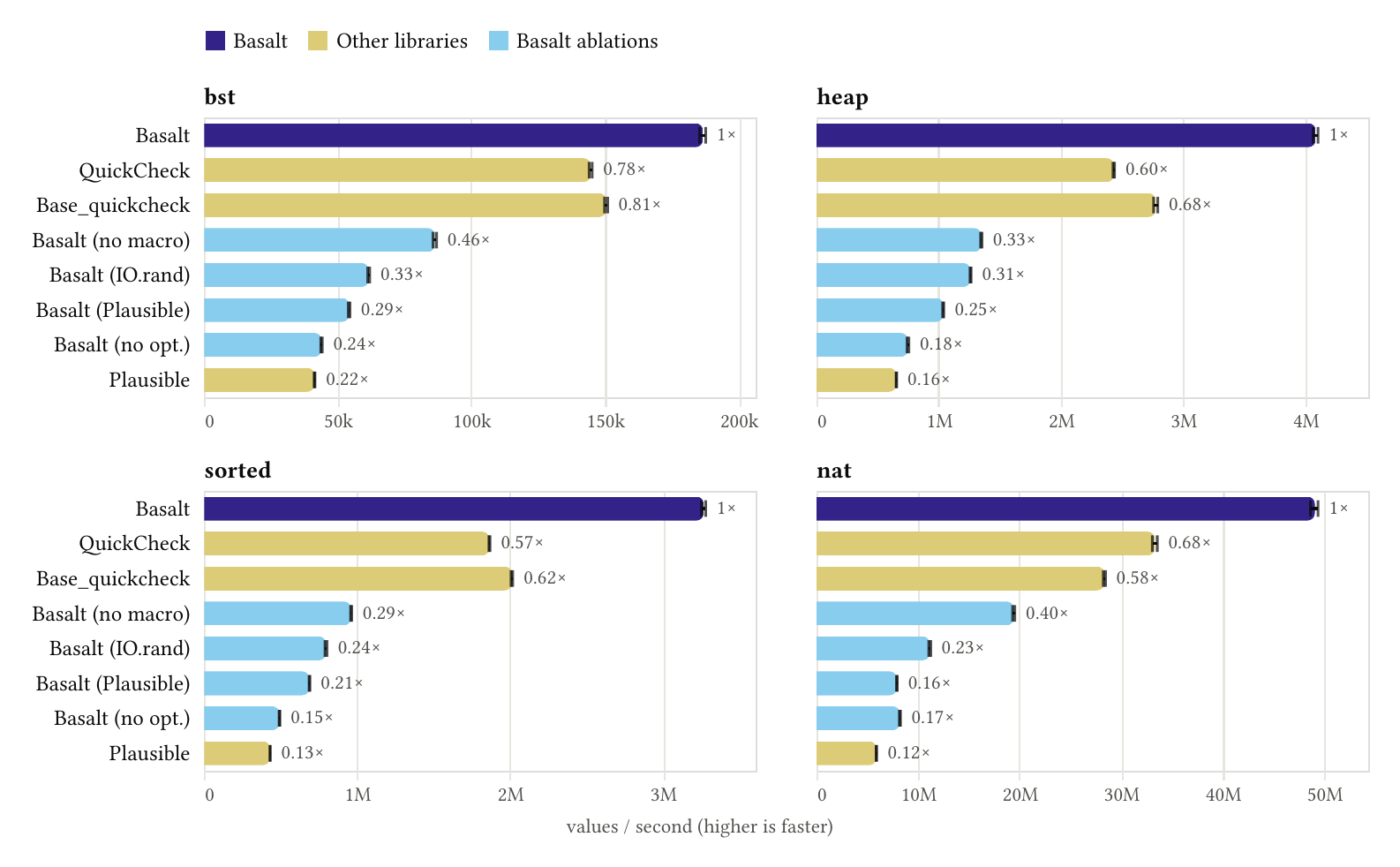}
\caption{Generation throughput of \ourlib vs. Haskell's QuickCheck vs. OCaml's Base\_quickcheck vs. Lean's Plausible.
We compared each implementation with \ourlib on each benchmark using Welch's t-test on log throughput.
We corrected the p-values for multiple comparisons with Holm's method.
Every difference is significant: all p-values are well below 0.001.
Experiments were completed on an M5 Max MacBook Pro.}\label{fig:io-ablation}
\Description{Four bar charts, displaying generator performance.}
\end{figure}

\autoref{fig:io-ablation} shows the results of a lightweight experiment that demonstrates \ourlib's performance.
It is faster than Haskell's QuickCheck, OCaml's Base\_quickcheck, and Lean's Plausible.
The experiment measures four simple workloads: valid binary search trees (BSTs), valid heaps, sorted lists, and unconstrained natural numbers.
We implemented each workload as idiomatically as possible while ensuring that each generator produces the same distribution of data.
\autoref{app:benchmark-generators} gives the generators for every implementation.
The figure shows generator throughput.
We chose the number of values to generate to ensure that start-up time does not dominate the measurement; a faster generator (e.g., for natural numbers) generates more samples (70M) than a slower generator (e.g., for BSTs, 280K).
The chart reports the geometric mean of the throughput over 20 trials.
We find that \ourlib is faster than the comparisons, and \emph{much} faster than \ourlib without the aforementioned enhancements.

\subsection{\ourlib as a Foundation for Automatic Generator Synthesis}
\label{sec:ports}

We designed \ourlib's generator API to match abstractions provided by popular PBT libraries, like QuickCheck.
Nevertheless, the need to hand-write generators is a known challenge to PBT adoption~\cite{goldsteinPropertyBasedTestingPractice2024}, so we also designed \ourlib as a potential backend for existing frameworks for automatic generator synthesis.
As a backend, \ourlib gives these frameworks' users the ability to write distributional proofs about their synthesized generators, and also allows the frameworks' developers to synthesize proofs about their synthesized generators.
To demonstrate this potential, we retrofitted two Lean generator synthesis tools~---~Palamedes~\cite{goldsteinSearchConstrainedRandom2026} and Specimen~\cite{StrataorgSpecimen2026} (based on automation in QuickChick~\cite{lampropoulosGeneratingGoodGenerators2017,paraskevopoulouComputingCorrectlyInductive2022})~---~to target \ourlib.

\paragraph{Palamedes}

Palamedes~\cite{goldsteinSearchConstrainedRandom2026} is a recently published library that uses deductive program synthesis to automatically produce PBT generators.
\ourlib and Palamedes use different generator abstractions (Palamedes uses a \emph{free generator}~\cite{goldsteinParsingRandomnessFree2022} as its core data structure), so the main task was swapping one for the other.
First, this means that some of the optimizations that Palamedes performs in the object language (by pattern-matching on the structure of a generator) need to be done in the meta-language (by pattern-matching on the syntax of the generator) when using \ourlib.
Second, the free generator structure approach means that Palamedes handles recursion implicitly, through the following constructor:
\begin{lean}
indexed : (Nat → Gen (Option α)) → Gen α
\end{lean}
The support of \li{.indexed f} is defined as the union of the supports of all \li{f n}, under the assumption that \li{f} is continuous.

The first of these issues turned out to be less of an issue than we thought, since Palamedes already needed to do some meta-programming to manage optimization under binders.
We converted a few object-language analyses and moved the rest to the meta-programming that Palamedes was already doing.
The bigger change was non-termination; rather than try to model \li{.indexed} directly, we opted to convert the recursion combinators (the \li{unfold} operations) to \li{partial_fixpoint}s and change the proof infrastructure to use our reasoning principles.

At the time of writing, Palamedes still only automates proofs for soundness and completeness (as it did in the original paper), but we think automating termination and cost proofs for a subset of generators should also be possible.
In the meantime, users can use AI-based programming tools to fill in the missing proofs.

\paragraph{Specimen}

Specimen~\cite{StrataorgSpecimen2026} is a Lean port of Rocq's QuickChick PBT library, which automatically
derives generators for data that must satisfy predicates expressed as inductive relations~\cite{elazarmittelmanTestingTheoremsFully2026}.
To derive such generators, Specimen attempts to find the best sub-generators for a relation's constructors.
Specimen finds these sub-generators by searching over candidate generators that instantiate a constructor's universally quantified variables and satisfy its hypotheses, all while maximizing a modular heuristic that approximates generator efficacy.
Specimen derives these hypothesis-satisfying
generators with a recursive call to the same search. For example, one might wish to derive a good generator 
for well-formed heaps satisfying this inductive relation:
\begin{lean}[aboveskip=2pt,belowskip=2pt]
inductive IsHeap : Nat → Tree Nat → Prop where
  | leaf : IsHeap lo leaf
  | node : lo ≤ x → IsHeap x l → IsHeap x r → IsHeap lo (node l x r)
\end{lean}
Running the command \li{derive_mutual generator (fun lo => ∃ t, IsHeap lo t)} derives a recursive generator of the same shape as
\li{genHeap}, which relies on a derived generator for \li{Nat}s \li{x} that satisfy \li{lo ≤ x} for some lower bound \li{lo}.

Modifying Specimen to emit \ourlib generators was straightforward because \emph{schedules}, Specimen's intermediate 
representation for the generators it derives, are already generic over the particular monad and combinators used
for each type of derivation.
Specimen's search results in a large dependency graph of these schedules, with some of those schedules possibly being defined 
mutually recursively; when producing generators, Specimen traverses this dependency graph and builds concrete programs.
Previously, Specimen's compiler expressed generators using the monad and combinators from the Lean PBT library, Plausible~\cite{leancommunityPlausible2025}; switching to \ourlib only required swapping Plausible's monad and generator combinators with \ourlib's combinators defined over an arbitrary \li{Generator} instance. 
These derived programs make use of sizes and backtracking, features \ourlib supports natively.

Once Specimen was underpinned with \ourlib, we were able to improve the library. 
Specimen is based on a new algorithm for generator synthesis, presented by \citet{elazarmittelmanTestingTheoremsFully2026}, and it does not support automatically derived proofs due to difficulties around mutually recursive derivations~\cite{StrataorgSpecimen2026}.
\ourlib helps address these issues, and our version of Specimen now supports proof automation.
By leveraging \ourlib's \li{walk} tactic and soundness and completeness infrastructure, alongside the proof strategy
from QuickChick's derived correctness proofs for non-mutually recursive generators~\cite{QuickChickRepo2026,paraskevopoulouComputingCorrectlyInductive2022,lampropoulosGeneratingGoodGenerators2017}, Specimen
is able to derive correctness proofs automatically for a wide range of generators. Soundness proofs use the same concise \ourlib proof script (\autoref{app:specimen-proofs}) for each mutually recursive group 
of generators. Completeness, which says that every term satisfying a relation has to be generable at a large enough size,
is proven by applying \li{walk} to each of the generator's branches corresponding to the constructors of the relation, examining whether the term could be generated by that constructor
and which choices could arrive at it.

Our experience re-engineering Palamedes and Specimen to emit \ourlib generators indicates that \ourlib's generator abstractions 
are sufficiently expressive to describe generators synthesized from existing tools, and that \ourlib's proof automation infrastructure 
can be effectively leveraged by existing synthesis tools.

\subsection{Case Studies}

Generators for binary search trees, min-heaps, and other data structures commonly presented in the PBT literature are small and well understood.
To show that \ourlib scales to realistic use-cases, which involve many non-trivial, interrelated data structures and validity predicates, we present two real-world case studies: Cedar, a domain-specific language used for writing authorization policies, and Strata, an intermediate verification language used for proofs about programs at a large technology company.
We find that in both cases, \ourlib was able to meet these projects' needs, and is now (or soon will be) in production use.

\subsubsection{Cedar Case Study}\label{sec:cedar}

Cedar is an open-source authorization policy language used by cloud services such as Amazon Verified Permissions.
The production implementation of Cedar is written in Rust, but, following an approach it calls
Verification-Guided Development~\cite{disselkoenHowWeBuilt2024}, the Cedar team maintains formally
verified models of key Cedar components in Lean~\cite{cedarspec2026}. To provide assurance that the production
implementation of Cedar behaves as expected, the two implementations of Cedar are differentially tested.
Performing this differential testing requires a good generator for Cedar policies. The Cedar team handwrote a
generator in Rust for Cedar policies. Writing this generator was expensive: according
to the Cedar team, it took about six months to get a working draft of the generator at about 5,400 lines of
code~\cite{elazarmittelmanTestingTheoremsFully2026}. The generator has required nonstop maintenance since then to track changes to Cedar, with 152 commits in
39 of the 41 months~\cite{cedarspec2026} since its initial release. To see if this generator maintenance was avoidable, 
we attempted to derive a generator for valid Cedar policies automatically using Specimen, and prove it correct using its new \ourlib backing.

Cedar's validator, which, given a Cedar schema and policy, decides whether the pair is valid, was initially written as an executable function, 
so to use Specimen on it we first had to translate it to an equivalent inductive relational specification. Specimen's derivation
creates a generator that inverts the relation to produce well-formed schemas and valid policies by construction.
To perform this translation, 
we used Claude Opus 5.0, iteratively refining the resulting relations with Specimen by deriving a generator and testing its output for 
divergence with the functional specification, before using the LLM to prove each relation's correspondence with its function.
The final specification describes Cedar's validation with 159 inductive relations.
Specimen uses these to derive generators for valid Cedar
expressions, policies, and schemas.
The dependency graph of generators
required to define
the top-level generators amounts to 291 definitions, about 54,500 lines of code (LoC), with 20 mutually recursive groups (\autoref{tab:cedar-derived}).

Re-engineering Specimen to emit \ourlib generators only results in slightly higher overhead when deriving generators (57 seconds vs.\ 49 seconds).
When the derived \ourlib generators are executed using the SplitMix PRNG (\autoref{sec:performance}), 
the generators run only 2\% to 3\% slower than when compiled directly to Plausible. Specimen is unable to take
advantage of the speed provided by \ourlib's macros as of yet. There is ample room for optimization. 
This experience demonstrates that \ourlib remains competitive with existing generator libraries and can be feasibly used as the 
foundational generator framework underpinning industrial PBT deployments.

\begin{table}[ht]
\centering
\small
\setlength{\tabcolsep}{4pt}
\caption{Overview of the Cedar generators derived using Specimen's \ourlib backend: no. of definitions, LoC, no. of mutually recursive groups in the dependency graph, and these groups' sizes. 
These figures are broken down across generators (Gen.), enumerators (Enum.), and checkers (Check.).
Enumerators and checkers support the generators derived by Specimen: enumerators are used to exhaustively search for values, e.g., when trying to validate an existential statement, and checkers help to filter invalid values from generators that are too permissive.}
\label{tab:cedar-derived}
\begin{tabular}{@{}lrrrrrrrrrr@{}}
\toprule
& \multicolumn{3}{c}{Definitions} & \multicolumn{3}{c}{Lines of derived code} & \multicolumn{4}{c}{Mutually recursive groups} \\
\cmidrule(lr){2-4}\cmidrule(lr){5-7}\cmidrule(l){8-11}
& Gen. & Enum. & Check. & Gen. & Enum. & Check. & Groups & Largest & Mean & Mean (${>}1$) \\
\midrule
Expressions &  95 & 138 & 11 & 24,425 & 25,352 & 1,060 & 181 & 21 & 1.35 & 4.32 \\
Policies    &  21 &   0 &  2 &  1,736 &      0 &    58 &  23 &  1 & 1.00 & --- \\
Schemas     &  24 &   0 &  0 &  1,886 &      0 &     0 &  23 &  2 & 1.04 & 2.00 \\
\midrule
Total       & 140 & 138 & 13 & 28,047 & 25,352 & 1,118 & 227 & 21 & 1.28 & 4.20 \\
\bottomrule
\end{tabular}
\end{table}

\paragraph{Proofs about the generated inputs}
Moving beyond the soundness and completeness proofs that Specimen now automates, we can explore interesting distributional properties.
Cedar's developers have repeatedly tuned their hand-written generators to control the size of generated inputs~\cite{cedarspec2026}.
We identified three recurring patterns in these tuning changes, and for each one we prove a corresponding bound on our derived generators (\autoref{app:cedar-theorems} states these bounds precisely).
Unlike soundness and completeness, which Specimen proves automatically, these bounds were proven using an LLM.

The first pattern concerns the number of request environments in a generated policy~\cite{cedarSpecIssue610,cedarSpecPR719,cedarSpecPR853}.
Cedar policies are validated against a \emph{schema}, which defines which principal and resource types each action applies to.
Each valid combination of principal type, action, and resource type is called a \emph{request environment}, and validation checks a policy once per request environment.
We prove that our derived schema generator respects a configurable cap on request environments and produces schemas with more than one request environment with at least 27\% probability (provided that such schemas are reachable based on the generator's arguments).

The second pattern involves entity hierarchies~\cite{cedarSpecPR1044,cedarSpecPR1051,cedarSpecPR1060,cedarSpecPR1062}.
Cedar policies refer to \emph{entities}: an application's principals, actions, and resources.
Entities can be grouped into hierarchies; for example, a photo may belong to several albums, and an album may itself belong to another album~\cite{cedarDocsEntities}.
Our derived schema generator supports a \emph{sparse} mode.
We prove that every hierarchy produced in this mode is a forest, and the expected number of ancestors of an entity type grows at most logarithmically in a configurable size parameter.

Finally, the Cedar team has repeatedly adjusted the depth of generated expressions~\cite{cedarSpecPR764,cedarSpecPR923,cedarSpecIssue942}.
A policy's conditions are Boolean expressions that can be nested arbitrarily deeply.
The depth of an expression cannot be bounded independently of its context: in a well-typed policy, an access \li{x.y.z} to an optional attribute \li{z} must appear under a check \li{x.y has z}, so an expression that uses the check must copy \li{x.y}.
We prove that the depth of every generated expression is bounded linearly in the product of the generator's two size parameters plus the maximum depth of any
checked attribute access it could contain (detailed in \autoref{app:cedar-theorems}).

\subsubsection{Strata Case Study}\label{sec:strata}

We also used \ourlib to develop generators for Strata~\cite{StrataorgStrata2026},
an intermediate verification language implemented in Lean.
Strata is under active use at a major technology company as part of a pipeline for
proving properties of programs in a variety of source languages, including Python.
Strata consists of a series of \emph{dialects} that represent programs
at different levels of abstraction, along with transformations between these dialects.
We used PBT to gain confidence in the correctness of those parts of Strata's implementation
whose proofs are still in development (e.g., semantics-preservation properties of certain transformations),
and of parts for which proofs are infeasible
(e.g., round-trip properties of Strata's \li{IO}-using pretty printer and parser).
To do this, we developed bespoke generators of well-typed programs in Strata's Core dialect using \ourlib.
In particular, we used Claude Opus 5.0 to automatically synthesize generators of
different Strata Core components and to prove properties of these generators.\footnote{
  All the Strata generators were developed from scratch using \ourlib, unlike the Cedar case study (\autoref{sec:cedar}), 
  where we compared \ourlib generators with existing ones.
}
We proved using \ourlib that the generators are sound and complete (up to some side conditions), and that they satisfy specific distributional properties.
The remainder of this subsection describes Strata in more detail, the generators and their properties, and the dozens of bugs we found using PBT.

\paragraph{Overview}
Strata's central dialect is dubbed
\emph{Strata Core} and takes inspiration from prior intermediate verification languages
such as Why3~\cite{bobot_et_al_2011} and Boogie~\cite{barnett_et_al_2006}.
Strata Core is stratified into several layers: the expression language is a $\lambda$-calculus
with Hindley--Milner-style prenex polymorphism, uninterpreted calls to
\emph{ambient functions} (built-in functions for arithmetic, string manipulation, etc.), and existential and universal quantification.
The imperative fragment includes mutable variables (including
non-deterministic assignment) and structured control flow (including loop
invariants). Top-level declarations include abstract types,
pure functions, stateful procedures with Dafny-style pre- and postconditions~\cite{leino_dafny_2010},
user-defined algebraic data types, and type aliases. 
Our generators support generating programs in a large subset of Strata Core (features not supported include recursive 
functions, and generating pre- and postconditions with a known verification outcome).

\begin{table}[t]
  \footnotesize
  \caption{Scale of the \ourlib generators for Strata Core and their proofs (left), along with the properties tested and bugs found (right).}
  \label{tab:strata-generators}
  \begin{subtable}[t]{0.61\textwidth}
    \centering
    \caption{Size of each Strata Core fragment's AST and its typing specification, and of our \ourlib generators and their proofs.
    (Constructor counts are summed across all AST data types.)}
    \label{table:strata-generators-size}
    \setlength{\tabcolsep}{2pt}
    \begin{tabular}{@{}lrrrr@{}}
      \toprule
      \textbf{Fragment}
        & \begin{tabular}[b]{@{}r@{}}\textbf{No.\ of}\\\textbf{constructors}\\\textbf{in Strata AST}\end{tabular}
        & \begin{tabular}[b]{@{}r@{}}\textbf{No.\ of}\\\textbf{typing rules}\end{tabular}
        & \begin{tabular}[b]{@{}r@{}}\textbf{No.\ of}\\\textbf{generator}\\\textbf{definitions}\end{tabular}
        & \begin{tabular}[b]{@{}r@{}}\textbf{No.\ of}\\\textbf{generator}\\\textbf{theorems}\\\textbf{\& lemmas}\end{tabular} \\
      \midrule
      Primitives                & --  & --  & 7   & 11    \\
      Types \& exprs.           & 21  & 10  & 30  & 363   \\
      Commands                  & 12  & 7   & 11  & 51    \\
      Statements                & 6   & 11  & 14  & 244   \\
      Procedures                & 12  & 3   & 3   & 66    \\
      Functions                 & 1   & 1   & 21  & 79    \\
      Data types                      & 2   & 21  & 15  & 137   \\
      Declarations              & 15  & 12  & 19  & 189   \\
      General purpose           & --  & --  & 3   & 45    \\
      \midrule
      \textbf{Total}            & \textbf{69} & \textbf{65} & \textbf{123} & \textbf{1,185} \\
      \bottomrule
    \end{tabular}
  \end{subtable}%
  \hfill
  \begin{subtable}[t]{0.37\textwidth}
    \centering
    \caption{Properties tested and bugs found in Strata via PBT (by category) using \ourlib generators. (Some spec bugs were found via testing, while others were found when proving generators correct.)}
    \label{table:strata_bugs}
    \setlength{\tabcolsep}{4pt}
    \begin{tabular}{@{}lrr@{}}
      \toprule
      \textbf{Category} & \begin{tabular}[b]{@{}r@{}}\textbf{Prop-}\\\textbf{erties}\end{tabular} & \begin{tabular}[b]{@{}r@{}}\textbf{Bugs}\\\textbf{found}\end{tabular} \\
      \midrule
      Parser                     & 1   &  2 \\
      Pretty-printer             & 6   & 10 \\
      Type checking \& inference & 17  &  3 \\
      Transformations            & 97  &  6 \\
      Symbolic execution         & 6   &  1 \\
      SMT encoding               & 8   &  4 \\
      Derived functions      & 3   &  2 \\
      Specifications             & --  &  9 \\
      PBT infrastructure         & 19  & -- \\
      \midrule
      \textbf{Total}             & \textbf{157} & \textbf{37} \\
      \bottomrule
    \end{tabular}
  \end{subtable}
\end{table}

Since Strata Core consists of multiple language layers,
our LLM-synthesized \ourlib generators likewise have a layered structure.
Both the generators and their correctness proofs compose to
produce results for full Strata Core programs.
For the expression layer, we generate well-typed terms in a type-directed manner, with the generator closely following the structure of typing rules,
as pioneered by \citet{palkaTestingOptimisingCompiler2011}.
For the imperative fragment of Strata Core, we generate commands and statements sequentially,
extending the typing context when new definitions are introduced, so that
later-generated portions can refer to earlier definitions.
Soundness and completeness proofs are performed by induction on the generator's \emph{size} parameter and case analysis on the
type of the generated term, following prior work
\cite{lampropoulosGeneratingGoodGenerators2017, paraskevopoulouComputingCorrectlyInductive2022}.

An initial version of these LLM-synthesized
proofs was done prior to the development of \ourlib's proof automation tactics (\autoref{sec:program-logics}).
These proofs were later refactored to use \ourlib's \leaninline{walk} tactic and auxiliary macros,
allowing for a cumulative reduction of over 4,000 lines of code across all generator proofs. 
For example, in the soundness proof for the expression generator, these proof automation tactics were able to 
reduce the proof size by approximately 90\%. In the original proofs, for each possible combination 
of \leaninline{size} and base type, the generator needs to choose among a list of sub-generators according to some weights,
requiring a proof that each choice of sub-generator produces a well-typed term. The original proof had to perform case analysis for each choice of sub-generator, then reason about the chosen sub-generator's support. 
This line of repetitive reasoning is automated by the \leaninline{walk} tactic, demonstrating how 
\ourlib's proof automation facilities effectively help LLMs synthesize more succinct proofs about generators.
\autoref{table:strata-generators-size} displays the size of the typing specification, generators and associated correctness proofs
for different fragments of Strata Core.

\paragraph{Controlling size of generated programs}
Beyond soundness and completeness, some properties we wanted to prove about our generators are lower bounds on the probability of generating nested statements (e.g., loops and conditionals),
and upper bounds on the expected frequency of such statements (the exact bounds may change based on generator tuning).%
\footnote{For testing loop-related optimizations, we have found it helpful to bias our generator toward programs that contain loops, and we wanted to see how frequently loops appear in generated programs.}
Proving these properties required changing our generators' design. 
Our initial generators used the same \leaninline{size} parameter to control the 
recursion depth and the maximum length of nested statement bodies. However, 
we realized that using this design, the best upper bound we could prove on the expected number of loops in generated programs grew factorially. 
We instead decided to decouple these parameters and limit the lengths of nested statements' bodies.
This change enabled us to prove bounds on the probability and expected frequency of AST nodes that are independent of \leaninline{size}. 
In practice, we choose generated instances to have a small \leaninline{size} to avoid potential blowups in generated programs.
However, reasoning about the parameterized distribution with \ourlib requires us
to handle these cases, making the generators more robust.

\paragraph{Bugs found}
With our generators, we regularly test over 150 properties, of which over 130 span the Strata codebase (the remaining properties are tests of 
the generators themselves and associated PBT infrastructure, e.g. QuickCheck-style shrinkers \cite{claessenQuickCheckLightweightTool2000}). For example, we test that the parser
and pretty-printer round-trip (\leaninline{parse(print(p)) = p} for all randomly generated programs \leaninline{p}),
and that common sub-expression elimination does not result in unbound variables. \autoref{table:strata_bugs} summarizes the Strata PBT effort,
conveying the number of properties tested and bugs found.
We now briefly discuss some notable bugs.
When testing the round-trip property for the parser and pretty-printer, we found that the polymorphic function \leaninline|foo<s>(x: s) { $\ldots$ }| (where \leaninline{s} is a type variable)
failed to parse, as \leaninline{<s} was also Strata's concrete syntax for the signed less-than operator over bit-vectors. This syntactic ambiguity
meant that users were unable to define polymorphic functions with type variables beginning with \leaninline{s}.
Another bug we found involves Strata's symbolic execution engine, which collects proof obligations expressed in the form
of assertions in the source program. We found that in programs containing two consecutive non-deterministic branches, proof obligations in one of the conditionals would be silently dropped.
This bug has since been fixed.
Bugs like these show how \ourlib generators helped uncover difficult-to-find bugs, thereby improving the reliability of Strata's codebase.

\section{Related Work}\label{sec:related}

\paragraph{Representing Generators for PBT}

The original QuickCheck paper~\cite{claessenQuickCheckLightweightTool2000} defines a PBT generator as a function from random seeds to values.
\citet{paraskevopoulouFoundationalPropertyBasedTesting2015} formalized the QuickCheck representation in Rocq, allowing users to prove correctness properties for generators.
They defined the \emph{support} of a generator~---~the set of values that it can potentially generate~---~as follows:
\begin{lean}
def QC.support (g : QC.Gen α) : Set α := {x | ∃ s, g s = x}
\end{lean}
In other words, a generator's support is the set of all values for which there exists a seed that produces the value.
The QuickChick representation is useful for many things, but it cannot capture proofs about distributions.
Our representation addresses this limitation, creating a stronger foundation for proofs about PBT moving forward.

\citet{goldsteinParsingRandomness2022} introduced \emph{free generators}, 
in which generators are represented as a data structure that can be interpreted differently in various contexts. Using this mechanism,
free generators can be interpreted as parsers for random sequences of choices. Our representation is inspired by free generators,
but we make generators interpretable both as executable functions and as proof objects, allowing users to directly prove properties 
about their generators' distributions. 
 
\citet{yu_et_al_2026} develop complexity-theoretic foundations for PBT generators, modeling generators as 
Turing machine transducers from random bits to outputs. They show that for certain classes of preconditions, 
efficient (polynomial time) constrained generators are impossible to achieve under standard complexity-theoretic assumptions, 
in which case one must settle for rejection sampling or restrict the generator's sample space.
\ourlib allows users to implement both constrained and rejection-sampling-based generators and prove distributional properties about them. 
Crucially, users are not limited to either generator design: they can 
switch between the two, based on what is theoretically possible and what meets their practical needs. 

\paragraph{Dynamic Alternatives for Constrained Generation}

Research across both the programming languages and fuzzing communities sometimes eschews static generators for dynamic approaches that constrain the generation space based on feedback gathered during testing.
These approaches often seem compelling, since writing generators like the ones we work with in \ourlib can be time-consuming.
But these dynamic approaches to generation have substantial downsides.
First, there is mounting evidence that stronger static generators are better at finding bugs than weaker generators enhanced by dynamic feedback.
\citet{vikramFuzzingAgentsGenerators2026} compared constrained generators that produce well-formed inputs by construction against generators that employ coverage-guided fuzzing and found that coverage guidance only contributes marginally to the branch coverage of generated inputs.
These results are corroborated by \citet{lampropoulosCoverageGuidedProperty2019}: they developed FuzzChick, a variant of Rocq's QuickChick framework which combines coverage guidance with automatically derived mutation functions for algebraic data types, interleaving coverage-guided mutation with random generation.
When evaluating FuzzChick against QuickChick, they found that bespoke generators that produce well-formed inputs by construction (e.g., QuickChick generators synthesized directly from an inductive relation~\cite{lampropoulosGeneratingGoodGenerators2017, paraskevopoulouComputingCorrectlyInductive2022}) remain more performant overall. 
Second, many dynamic approaches to generation do not guarantee distributional properties like the ones \ourlib can prove.
Coverage-guided fuzzing~\cite{godefroidGrammarbasedWhiteboxFuzzing2008}, including \emph{parametric generator-based fuzzing}~\cite{padhyeJQFCoverageguidedPropertybased2019,padhyeSemanticFuzzingZest2019,crowbar}, makes no claims to approximate any particular distribution.
Indeed, existing work has shown that some parametric fuzzing schemes are subject to the \emph{havoc paradox}~\cite{li2025havoc}, in which changing the randomness for a generator has unpredictable impacts on the distribution of the generated values.
Solver-based approaches~\cite{deweyAutomatedDataStructure2015,steinhofelInputInvariants2022,seidelTypeTargetedTesting2015} are similarly limited.

Nevertheless, it would be interesting to see how \ourlib might interact with some limited kinds of dynamic feedback.
In \ourlib, since generators are implemented as type classes that can be interpreted differently for various use-cases, one could envisage developing a ``coverage-guided fuzzing'' instance of the \leaninline{Generator} type class, which allows users to still prove correctness and distributional properties of their generators, but also reap the benefits of coverage guidance.
Our hope is that one could prove some properties about the implementation once-and-for-all, e.g., that a particular fuzzing scheme avoids the havoc paradox.

\paragraph{SPMFs as a Semantic Domain}

\ourlib follows a long tradition in the probabilistic programming languages literature.
Many languages over the years have used SPMFs or a similar structure to represent potentially diverging distributions, including seminal work by \citet{kozenSemanticsProbabilisticPrograms1979} and \citet{jonesProbabilisticPowerdomainEvaluations1989} and recent work by \citet{zilbersteinDemonicOutcomeLogic2025} and \citet{hetzenbergerCostAwareProbabilityMonad2026}.
While \ourlib is the first application of SPMFs to PBT that we know of, we are certainly not the first to apply them to distributional semantics.
Our SPMF monad is also based on the Giry monad~\cite{giryCategoricalApproachProbability1982} for discrete distributions, which has been replicated before in various forms.

SPMFs also arise in other libraries for theorem proving.
For example, CryptHOL~\cite{lochbihlerProbabilisticFunctionsCryptographic2016} uses a shallowly embedded SPMF monad to capture and prove properties of cryptographic systems in Isabelle/HOL, and ALEA~\cite{audebaudProofsRandomizedAlgorithms2009} has a similar scheme for proving properties of randomized algorithms in Rocq.
In Lean, VCVio~\cite{cryptoeprint:2026/899} provides an SPMF-based semantics as a basis for a program logic.
\ourlib's contribution is to apply these ideas to PBT.

\section{Future Work and Conclusion}\label{sec:conclusion}

We have presented \ourlib, a foundational library for PBT generators with a probabilistic semantics.
In \ourlib, the generator code that runs for testing (at least as fast as generators in established PBT frameworks) is the same code that supports proofs about soundness, completeness, distributions, termination, and cost.
Because \ourlib represents structurally non-terminating generators without fuel, natural generators like rejection samplers keep the distributions their authors intended.
This foundation is strong enough to support automation: Palamedes and Specimen now produce \ourlib generators, and Specimen derives correctness proofs for the generators it builds.
In industrial case studies, \ourlib generators for Cedar were proved sound, complete, and within the size limits its developers asked for, and \ourlib generators helped find bugs in Strata.
Looking ahead, we see two directions for building on this foundation.

\paragraph{Using Proofs as Tuning Feedback}
\ourlib lets programmers prove that a generator's distribution falls short of a specification, but it does not help fix the generator.
Existing work~\cite{tjoaTuningRandomGenerators2025,zhuangAriadneAutomaticallyTuning2026} has begun to explore automated strategies for tuning PBT generators, and \ourlib could complement those tools well.
In particular, we think that counterexamples from failing proof automation could be useful feedback for automated tuning systems.
We hope these approaches can close the loop on high-quality generator automation: a developer will be able to write down a generator in \ourlib (or get one from Palamedes or Specimen), express some distributional goal, and automatically obtain a generator that provably meets that goal.

\paragraph{Building a Library of High-Quality Generators}
\ourlib also gives the PBT community a common language for sharing generators.
We envision a library that supplies generators from a wide range of domains, alongside proofs that guarantee the quality of those generators.
If a programmer needs a generator for SQL queries, PNG files, or simple Python programs, they may be able to take one off the shelf, read through its theorems, and run it against their codebase.
For more bespoke tasks, programmers can use existing library generators as a starting point and build their own (which, hopefully, they will contribute back to the library when they are done).

\section*{AI Use}

AI was used as a programming tool in this work; it was used while designing \ourlib itself, while building our case study implementations (especially where mentioned above), and while setting up our experiments.
All AI-generated code was extensively reviewed by humans.
AI was also used in small ways for editing tasks in the paper itself (e.g., moving a section to the appendix or checking for typos and inconsistencies), but no text in the paper was written by AI.

\bibliographystyle{ACM-Reference-Format}
\bibliography{references}

\appendix

\section{Primitives for Randomness}\label{app:primitives-for-randomness}

The \leaninline{choose} operator was not our only available option for a primitive randomness operator, but it has a number of qualities that make it work well for our use-cases.
First, it is relatively efficient; primitives like \leaninline{flip} and \leaninline{coin}, which produce only one bit of randomness at a time, are theoretically nice but impractical to implement.
Second, it is first-order.
Some formalisms make combinators primitive, like \li{oneOf}, which makes a uniform random choice between a list of generators, or \li{frequency}, which makes a weighted choice between sub-generators, but both of these are higher-order and can be theoretically challenging as a built-in combinator.

Our \leaninline{choose} also has some disadvantages.
Using \leaninline{ℕ} ranges (as opposed to \leaninline{UInt64}) makes proofs much easier, but it means that implementations either need to sample integers of arbitrary size or else diverge slightly from the true semantics.
We have never seen this cause real issues, but it is a small bit of friction between theory and practice.
Also, any single choice of randomness primitive locks the implementation into that granularity of randomness.
If another option becomes available, it is difficult to adapt the existing formalism to that approach.
This cost is unavoidable when building a static formalism.

\section{Proving Faithfulness}\label{app:proving-faithfulness}

Connecting the \li{IO} interpretation to the \li{SPMF} interpretation requires a series of steps, each of which we validate in different ways.
\begin{enumerate}\setlength{\itemsep}{\medskipamount}
  \item The \li{IO} interpretation agrees with an ``\li{IOModel}'' interpretation that abstracts the C code in Lean. \\
  \textit{Challenge:} The \li{IO} interpretation calls C code on a global PRNG state, whereas the model is a pure Lean interpretation of SplitMix. \\
  \textit{Validation:} We run both implementations of SplitMix on 1,000 random seeds, and ensure that each produces exactly the same stream of 100 values.
  We also require that generators be run in a single-threaded environment.
  \item The ``\li{IOModel}'' interpretation agrees with an ``\li{IdealizedModel}'' interpretation that is assumed to be a perfect PRNG. \\
  \textit{Challenge:} The idealized model assumes that draws from the PRNG are independent and uniformly random. \\
  \textit{Validation:} While this is strictly speaking false of any PRNG, SplitMix is known to be a good approximation of a real RNG. (Note: QuickChick also makes this assumption.)
  \item The ``\li{IdealizedModel}'' interpretation agrees with the \li{SPMF} interpretation. \\
  \textit{Challenge:} This fact is easy to prove of each \li{Generator} operation, but the end-to-end argument requires parametricity. \\
  \textit{Validation:} We prove this fact on a per-generator basis, using a logical relation that connects the two interpretations. The \li{walk} tactic allows us to fully automate these proofs for users.
\end{enumerate}
In the end, this chain of reasoning shows that the \li{IO} interpretation samples from a reasonable approximation of the \li{SPMF} distribution.

\section{Size}\label{app:size}

\ourlib's \li{Generator} class and its standard interpretations do not include a size as a primitive, but they support running and reasoning about generators that rely on a size parameter.
\ourlib manages size with an additional type class that carries the standard QuickCheck size operations:
\begin{lean}
class Sized (g : Type → Type) where
  sized : (Nat → g α) → g α
  resize : Nat → g α → g α
\end{lean}
And we can add sizes to any generator by just adding a \li{Nat} argument:
\begin{lean}
def WithSize (G : Type → Type v) (α : Type) : Type := Nat → G α
instance [Generator G] : Generator (WithSize G) where ...
instance [Monad G] : Sized (WithSize G) where ...
\end{lean}
If a user wants to use size combinators in their generator, they can simply add a \li{Sized} constraint alongside the \li{Generator} constraint.
For example, this is how Haskell's QuickCheck defines the \li{listOf} combinator:
\begin{lean}
def listOf [Generator G] [Sized G] (gen : G α) : G (List α) :=
  sized fun n => do
    let k ← chooseNat 0 n
    vectorOf k gen
\end{lean}
When the user finally interprets a generator like this, the size falls out naturally as an argument.
For proofs, the user can interpret this generator at type \leaninline{Nat → SPMF (List α)} (instantiating \leaninline{G} as \leaninline{WithSize SPMF}), at which point they can decide whether to supply a concrete size or quantify over sizes. For execution, the user can interpret at \leaninline{Nat → IO (List α)} (instantiating \leaninline{G} as \leaninline{WithSize IO}) and sample at whatever size they feel is appropriate.

\section{Failure and Backtracking}\label{app:backtracking}

Much like with \li{Sized}, we can also layer failure and backtracking into generators via auxiliary monads.
If a generator should be allowed to fail, we can give it type \leaninline{∀ [Generator G], OptionT G α}; for example:
\begin{lean}
def tryGenFive [Generator G] (tries : Nat) : OptionT G Nat := do
  match tries with
  | 0 => failure
  | tries' + 1 =>
    let x ← genNat
    if x == 5 then return x else tryGenFive tries'
\end{lean}
This works out because \li{OptionT G} can be made an instance of \li{Generator} for any \li{G} that is itself a \li{Generator}.
We build some lightweight proof infrastructure to help users deal with generators like this, but most of the standard \ourlib infrastructure continues to work with partial generators.

\Needspace{20\baselineskip}
\section{A Fuel-Based Heap Generator}\label{app:fuel}

In QuickChick, the \li{genHeap} generator from \autoref{fig:genNat-genHeap} would need to look something like the following.
The code that manages fuel is highlighted in yellow.
\begin{lean}[style=leandiff]
def genHeapStruct [Generator G] @(fuel : Nat)@ (lo : Nat) : @OptionT@ G (Tree Nat) :=
  @match fuel with@
  @| 0 => failure@
  @| fuel' + 1 =>@
    oneOf [
      fun () => return leaf,
      fun () => do
        let delta ← genNat
        let x := lo + delta
        let l ← genHeapStruct @fuel'@ x
        let r ← genHeapStruct @fuel'@ x
        return (node l x r)
    ]
\end{lean}

\section{Proofs of Generator Properties}\label{app:proofs}

This appendix gives the full proofs of the theorems stated in \autoref{sec:proofs}.
Many of them use \li{walk}, the custom \ourlib tactic described in \autoref{sec:program-logics}.

\subsection{Soundness and completeness of \texorpdfstring{\li{genHeap}}{genHeap}}
\begin{lean}
theorem genHeap.sound_complete :
    IsSoundAndComplete (genHeap lo) (isHeap lo) := by
  refine .intro ?sound ?complete
  case sound =>
    rw [IsSoundFor.iff_obs]
    walk fixpoint [genNat.sound_complete.sound.obs]
    all_goals simp_all [isHeap]
  case complete =>
    refine IsCompleteFor.of_measure
      (gen := fun lo => (genHeap lo : SPMF (Tree Nat)))
      (fun _ t => t.size) (fun n ih lo t hn h => ?_) lo
    rw [genHeap, SPMF.mem_support_iff_may]
    walk [genNat.sound_complete.complete.obs, SPMF.le_may_of_measure ih]
    cases t with
    | leaf => simp
    | node l x r =>
      obtain ⟨d, rfl⟩ := Nat.exists_eq_add_of_le h.1
      grind [= size, = isHeap]
\end{lean}

\subsection{Probability that \texorpdfstring{\li{genHeap}}{genHeap} returns the empty heap}
\begin{lean}
theorem genHeap.prob_empty : ProbabilityOf (genHeap lo) leaf = 1 / 2 := by
  rw [genHeap]
  simp [SPMF.prob, SPMF.expect_oneOf, SPMF.expect_bind, SPMF.expect_map,
    SPMF.expect_const, Function.comp_def]
\end{lean}

\subsection{Expected value of \texorpdfstring{\li{genNat}}{genNat}}
\begin{lean}
theorem genNat.expect_le : ExpectedValue genNat (fun n => n) ≤ 1 := by
  rw [SPMF.expect_eq_obs]
  walk fixpoint
  norm_num
\end{lean}

\subsection{Almost-sure termination of \texorpdfstring{\li{genNat}}{genNat}}
\begin{lean}
theorem genNat.terminates : IsAlmostSurelyTerminating genNat := by
  mass_fixpoint using SPMF.LfpIsOne.affine (m := 1 / 2) (by norm_num)
  simp [ENNReal.div_eq_inv_mul, mul_add]
\end{lean}

\subsection{Cost bound for \texorpdfstring{\li{genHeap}}{genHeap}}
\begin{lean}
theorem genHeap.cost_bounded :
    IsCostBounded (genHeap lo) (fun t => 3 * t.size + t.sum + 1) := by
  rw [IsCostBounded.iff_obs]
  walk fixpoint [genNat.cost_bounded.obs]
  all_goals simp only [Tree.size, Tree.sum]; omega
\end{lean}

\subsection{Expected cost bound for \texorpdfstring{\li{genNat}}{genNat}}
\begin{lean}
theorem genNat.expected_cost_bounded :
    IsExpectedCostBounded genNat 2 := by
  rw [IsExpectedCostBounded.iff_obs]
  walk fixpoint
  exact ENNReal.div_le_of_le_mul (by norm_num)
\end{lean}

\subsection{No expected cost bound for \texorpdfstring{\li{genHeap}}{genHeap}}
\begin{lean}
theorem genHeap.not_expected_cost_bounded {B : ℝ≥0∞} (hB : B ≠ ⊤) (lo : Nat) :
    ¬ IsExpectedCostBounded (genHeap lo : SPMF.Cost (Tree Nat)) B := by
  refine IsExpectedCostBounded.not_of_step_iInf
    (g := fun lo => genHeap lo) one_ne_zero hB (fun lo => ?_) lo
  have hm := fun lo' =>
    (genHeap.cost_faithful (lo := lo')).one_le_mass genHeap.terminates
  have hnat := genNat.cost_faithful.one_le_mass genNat.terminates
  conv_rhs => rw [genHeap]
  walk [fun lo' => SPMF.Cost.le_expect_add (hm lo')
      (iInf_le (fun j => SPMF.Cost.expectObs.spec (genHeap j : SPMF.Cost (Tree Nat))
        (fun _ n => (n : ℝ≥0∞))) lo'),
    SPMF.Cost.le_expect_add (x := genNat) (h := fun _ n => (n : ℝ≥0∞))
      hnat zero_le]
  generalize ⨅ j, SPMF.Cost.expectObs.spec (genHeap j : SPMF.Cost (Tree Nat))
    (fun _ n => (n : ℝ≥0∞)) = m
  norm_num
  rw [ENNReal.le_div_iff_mul_le (by norm_num) (by norm_num)]
  exact le_of_eq (by ring)
\end{lean}

\section{An Example of the \texorpdfstring{\li{walk}}{walk} Tactic}\label{app:walk-example}

Consider proving the following theorem:
\begin{lean}
theorem genPairFiveToTen.sound :
    IsSound genPairFiveToTen (fun (x, y) => x ≤ 20 ∧ y ≤ 20)
\end{lean}
After applying the \li{walk} tactic, we are left with two goals:
\begin{lean}
(x : ℕ), (hx : 5 ≤ x ∧ x ≤ 10) ⊢ x ≤ 20
(y : ℕ), (hy : 5 ≤ y ∧ y ≤ 10) ⊢ y ≤ 20
\end{lean}
Each of which can be easily discharged by \li{grind} or similar.

\section{Specimen's Soundness Proofs}\label{app:specimen-proofs}

Soundness for a Specimen generator states that for each relation \li{R} and input
arguments \li{ins} a successful generation of output arguments \li{outs} from its derived generator \li{genR}
must satisfy \li{R ins outs}.

\begin{lean}
theorem genR.sound : ∀ fuel initSize size ins,
    succObs.spec (genR fuel initSize size ins) (fun outs => R ins outs) :=
  by intro fuel
    induction fuel with
    | zero => intros; rw [genR.eq_def]; walk
    | succ f ih =>
      intros; rw [genR.eq_def]
      unfold_gen_instances [ih, facts]
      walk [ih, facts]
      all_goals ((try check_hyps [facts]); ctor_leaf)
\end{lean}

Though not required by \ourlib generators, Specimen generators rely on structural termination 
through a fuel argument, which the proof inducts on. At fuel \li{zero}, the generator fails, 
which is irrelevant to the specification \li{succObs}, which only focuses on values 
that the generator produces successfully. For non-zero fuel, the inductive hypothesis
gives access to soundness proofs for successful recursive calls. Specimen generators construct data that 
satisfies their relations by mirroring each rule of \li{R} with a branch of the generator satisfying it, so that
a successful generation of a branch amounts to an implicit proof that \li{R} holds. By way of \ourlib's \li{walk} tactic and \li{check_hyps} (a Specimen tactic),
the proof script uses the inductive hypothesis, \li{ih}, to convert the recursive generator calls to proofs of inductive uses of \li{R} 
in the constructor, and \li{facts}, the soundness proofs about generator dependencies, 
to convert non-recursive calls into proofs of the other relations that constructor relies on. For mutually recursive
groups, the \li{ih} is the conjunction of the soundness statements for all the group members. Finally,
\li{ctor_leaf} (a Specimen tactic) applies \li{R}'s constructor to the proofs of the inductive relations required to satisfy it reflected
from the generator calls.

\section{Theorems for the Cedar Generators}\label{app:cedar-theorems}

This section states the size and generator properties summarized in \autoref{sec:cedar}.
Specimen generators are parameterized by at least two sizes: $n$ indicates the maximum 
recursion depth of the generator and $\ell$ indicates the maximum size of any data generated
as an arbitrary member of a type (like a random string or integer). These properties are defined 
with respect to these two sizes, and for the schema generator, a budget $b$ intended to cap the
number of request environments its schemas have.

\begin{enumerate}
  \item \textsc{Support.}~Regardless of whether a schema is generated with a sparse, flat, or dense hierarchy graph,
  the number of request environments the generated schema has is capped by the configurable budget $b$.
  \begin{lean}
  theorem schema_combinations (b n l : Nat) (m : HierMode) (Γ : TypeEnv)
    (h : (m, Γ) ∈ supp (schemaGen b n l)) :
    Γ.schema.environments.length ≤ b
  \end{lean}
  \item \textsc{Distribution.}~With a budget of at least one, the schema generator always succeeds
  at generating a schema. 
  \begin{lean}
  theorem schema_never_fails (b n l : Nat) (hb : 1 ≤ b) : 
    Pr[schemaGen b n l succeeds] = 1
  \end{lean}
  \item \textsc{Distribution.}~With a budget of at least 2 and $\ell \geq 1$, the probability of a schema with
  at least 2 entity types having at least 2 request environments is at least $\frac{5}{18}$.
  \begin{lean}
  theorem schema_varied (b n l : Nat) (hb : 2 ≤ b) (hl : 1 ≤ l) :
    Pr[(_, Γ) ← schemaGen b n l : 2 ≤ Γ.schema.environments.length 
      | 2 ≤ Γ.ets.toList.length]
      ≥ 5 / 18
  \end{lean}
  \item \textsc{Support.}~The schema generator can choose to generate entity hierarchies of different
  densities. When it chooses to generate in the \li{sparse} mode, it generates a forest.
  \begin{lean}
  theorem sparse_hierarchy_forest (b n l : Nat) (Γ : TypeEnv) 
      (h : (.sparse, Γ) ∈ supp (schemaGen b n l)) :
      ∀ t ∈ hierarchy Γ, t.2 = [] ∨ ∃ p ps, t.2 = p :: ps ∧ (p, ps) ∈ hierarchy Γ
  \end{lean}
  \item \textsc{Distribution.}~When the schema generator succeeds in the \li{sparse} mode and the schema contains
  at least one standard entity type (\li{hierarchy Γ ≠ []}), the average number of ancestors of entity types 
  in the hierarchy has expectation upper bounded by $\ln(\ell + 1)$.
  \begin{lean}
  theorem sparse_mean_ancestors (b n l : Nat) :
      𝔼[(m, Γ) ← schemaGen b n l : .ofReal (meanAncestors Γ) 
        | m = .sparse ∧ hierarchy Γ ≠ []]
        ≤ .ofReal (Real.log (l + 1))
  \end{lean}
  \item \textsc{Support.}~For any set of capabilities, \li{c}, which encode the enclosing checks whose expressions
  may be copied into a generated expression and whose depth is not limited by $n$ or $\ell$, any type environment \leaninline|Γ|, type \leaninline|τ|, the
  well-typed expression generator at those capabilities, environment, and desired type only produces
  expressions \li{e} whose depth is bounded by the sum of the depth of the enclosing checks and a term linear 
  in the product of $n$ and $\ell$.
  \begin{lean}
  theorem expr_depth (decl syn : Bool) (c : Capabilities) (Γ : TypeEnv) 
                   (τ : CedarType) (n l : Nat) (e : Expr) (c' : Capabilities) 
    (h : (e, c') ∈ supp (exprGen decl syn c Γ τ n l)) :
    exprDepth e ≤ n * (l + 2) + capsDepth c
  \end{lean}
\end{enumerate}

\section{Micro-Benchmark Generators}\label{app:benchmark-generators}

The following generators are used for the micro-benchmarks in \autoref{sec:performance}.
There are four workloads:
\begin{itemize}
  \item \textbf{nat} generates an unconstrained natural number. It is \li{genNat} from \autoref{fig:genNat-genHeap}.
  \item \textbf{bst} generates a valid binary search tree with keys from 0 to 999.
  \item \textbf{heap} generates a valid min-heap. It is \li{genHeap} from \autoref{fig:genNat-genHeap}, with weights 5:4 instead of 1:1.
  \item \textbf{sorted} generates a sorted list of natural numbers, with weights 1:9.
\end{itemize}
We changed the weights for \textbf{heap} and \textbf{sorted} to make the distributions more interesting.
The expected sizes are 1 for \textbf{nat}, about 137.3 nodes for \textbf{bst}, 4 nodes for \textbf{heap}, and 9 elements for \textbf{sorted}.
Every implementation generates the same four distributions.

\subsection{\ourlib}\label{app:bench-basalt}

The two tree types are ordinary inductive types:
\begin{lean}
inductive BST.Tree (α : Type) where
  | leaf : BST.Tree α
  | node : BST.Tree α → α → BST.Tree α → BST.Tree α

inductive Heap.Tree where
  | leaf : Heap.Tree
  | node : Heap.Tree → Nat → Heap.Tree → Heap.Tree
\end{lean}
The generators use the \li{oneOf!} and \li{frequency!} macros, which expand to a chain of \li{if}s on literal running totals.
\begin{lean}
def genNat [Generator G] : G Nat := do
  oneOf! [
    fun _ => pure 0,
    fun _ => do
      let n ← genNat
      pure (n + 1)
  ]
partial_fixpoint
\end{lean}
\begin{lean}
def genWeightedBST [Generator G] (lo hi : Int) : G (BST.Tree Int) := do
  if h : lo > hi then
    return .leaf
  else
    frequency! [
      (1, fun _ => pure .leaf),
      (5, fun _ => do
        let x ← chooseInt lo hi (by omega)
        let l ← genWeightedBST lo (x - 1)
        let r ← genWeightedBST (x + 1) hi
        return .node l x r)
    ] (by simp)
partial_fixpoint
\end{lean}
\begin{lean}
def genHeap [Generator G] (lo : Nat) : G Heap.Tree :=
  frequency! [
    (5, fun _ => pure .leaf),
    (4, fun _ => do
      let delta ← genNat
      let x := lo + delta
      let l ← genHeap x
      let r ← genHeap x
      return .node l x r)
  ] (by simp)
partial_fixpoint
\end{lean}
\begin{lean}
def genSortedGt [Generator G] (m : Nat) : G (List Nat) := do
  frequency! [
    (1, fun _ => pure []),
    (9, fun _ => do
      let delta ← genNat
      let x := m + delta
      let xs ← genSortedGt x
      return x :: xs)
  ] (by simp)
partial_fixpoint
\end{lean}
The ``no macro'' implementation uses the same code with \li{oneOf!} and \li{frequency!} replaced by the functions \li{oneOf} and \li{frequency}.
The ``\li{IO.rand}'' implementation uses the standard IO PRNG instead of SplitMix.
The ``no opt.''\ implementation does both.

\subsection{Plausible}\label{app:bench-plausible}

These generators are written the way a Plausible user would write them: as \li{partial def}s at Plausible's \li{Gen} monad, using Plausible's \li{Gen.oneOf} and \li{Gen.frequency}.
They reuse the tree types from \autoref{app:bench-basalt}.
\begin{lean}
partial def genNat : Gen Nat :=
  Gen.oneOf #[
    pure 0,
    do
      let n ← genNat
      pure (n + 1)
  ] (by simp)
\end{lean}
\begin{lean}
partial def genWeightedBST (lo hi : Int) : Gen (BST.Tree Int) :=
  if h : lo > hi then
    pure .leaf
  else
    Gen.frequency (pure .leaf) [
      (1, pure .leaf),
      (5, do
        let ⟨x, _⟩ ← Gen.choose Int lo hi (by omega)
        let l ← genWeightedBST lo (x - 1)
        let r ← genWeightedBST (x + 1) hi
        return .node l x r)
    ]
\end{lean}
\begin{lean}
partial def genHeap (lo : Nat) : Gen Heap.Tree :=
  Gen.frequency (pure .leaf) [
    (5, pure .leaf),
    (4, do
      let delta ← genNat
      let x := lo + delta
      let l ← genHeap x
      let r ← genHeap x
      return .node l x r)
  ]
\end{lean}
\begin{lean}
partial def genSortedGt (m : Nat) : Gen (List Nat) :=
  Gen.frequency (pure []) [
    (1, pure []),
    (9, do
      let delta ← genNat
      let x := m + delta
      let xs ← genSortedGt x
      return x :: xs)
  ]
\end{lean}

\subsection{QuickCheck}\label{app:bench-quickcheck}

We use QuickCheck 2.19, compiled with GHC 9.10.3 and \texttt{-O2}.
The tree and list types have strict fields, so every value is fully built when it is generated.
\begin{lean}[language=Haskell,mathescape=false]
data Tree = Leaf | Node !Tree !Int !Tree
data SList = Nil | Cons !Int !SList
\end{lean}
\begin{lean}[language=Haskell,mathescape=false]
genNat :: Gen Int
genNat = oneof [pure 0, do n <- genNat; pure $! n + 1]
\end{lean}
\begin{lean}[language=Haskell,mathescape=false]
genBST :: Int -> Int -> Gen Tree
genBST lo hi
  | lo > hi = pure Leaf
  | otherwise =
      frequency
        [ (1, pure Leaf)
        , (5, do
            x <- chooseInt (lo, hi)
            l <- genBST lo (x - 1)
            r <- genBST (x + 1) hi
            pure (Node l x r))
        ]
\end{lean}
\begin{lean}[language=Haskell,mathescape=false]
genHeap :: Int -> Gen Tree
genHeap lo =
  frequency
    [ (5, pure Leaf)
    , (4, do
        delta <- genNat
        let !x = lo + delta
        l <- genHeap x
        r <- genHeap x
        pure (Node l x r))
    ]
\end{lean}
\begin{lean}[language=Haskell,mathescape=false]
genSorted :: Int -> Gen SList
genSorted m =
  frequency
    [ (1, pure Nil)
    , (9, do
        delta <- genNat
        let !x = m + delta
        xs <- genSorted x
        pure (Cons x xs))
    ]
\end{lean}

\subsection{Base\_quickcheck}\label{app:bench-base-quickcheck}

We use Base\_quickcheck v0.17.1, compiled with OCaml 5.4.1, flambda, and \texttt{-O3}.
Base\_quickcheck's own \li{weighted_union} draws a float, so we wrote \li{frequency} by hand on \li{int_uniform_inclusive}.
\begin{lean}[language={[Objective]Caml},mathescape=false]
let frequency (gs : (int * 'a Generator.t) list) : 'a Generator.t =
  let total = List.fold_left (fun acc (w, _) -> acc + w) 0 gs in
  let rec pick n = function
    | [] -> assert false
    | (k, g) :: gs -> if n < k then g else pick (n - k) gs
  in
  let
  pick n gs
\end{lean}
\begin{lean}[language={[Objective]Caml},mathescape=false]
type bst = Leaf | Node of bst * int * bst

let nat : int Generator.t =
  Generator.fixed_point (fun nat ->
    Generator.union
      [ return 0
      ; (let
\end{lean}
\begin{lean}[language={[Objective]Caml},mathescape=false]
let rec gen_bst lo hi : bst Generator.t =
  if lo > hi
  then return Leaf
  else
    frequency
      [ 1, return Leaf
      ; ( 5
        , let
          let
          let
          return (Node (l, x, r)) ) ]
\end{lean}
\begin{lean}[language={[Objective]Caml},mathescape=false]
let rec gen_heap lo : bst Generator.t =
  frequency
    [ 5, return Leaf
    ; ( 4
      , let
        let x = lo + delta in
        let
        let
        return (Node (l, x, r)) ) ]
\end{lean}
\begin{lean}[language={[Objective]Caml},mathescape=false]
let rec gen_sorted m : int list Generator.t =
  frequency
    [ 1, return []
    ; ( 9
      , let
        let x = m + delta in
        let
        return (x :: xs) ) ]
\end{lean}

\end{document}

%% file: preamble.tex
\newif\iflater\latertrue
\newif\ifdraft\draftfalse
\newif\ifafterdeadline\afterdeadlinefalse

\setcopyright{none}

\acmConference[PLDI '27]{ACM SIGPLAN Conference on Programming Language
  Design and Implementation}{June 2027}{TBD}
\acmYear{2027}
\acmISBN{}
\acmDOI{}

\usepackage{leanlst}
\usepackage{xspace}
\usepackage{epigraph}
\usepackage{wrapfig}
\usepackage{mathpartir}
\usepackage{booktabs}
\usepackage{array}
\usepackage{subcaption}
\usepackage{needspace}

\makeatletter
\newcommand{\@autorefsection}{\S\kern-\fontdimen2\font}
\newcommand{\setautorefname}[2]{%
  \expandafter\def\csname #1autorefname\endcsname{#2}}
\AtBeginDocument{%
  \setautorefname{section}{\@autorefsection}
  \setautorefname{subsection}{\@autorefsection}%
  \setautorefname{subsubsection}{\@autorefsection}%
  \setautorefname{paragraph}{\@autorefsection}%
  \setautorefname{subparagraph}{\@autorefsection}%
  \setautorefname{appendix}{Appendix}%
  \setautorefname{figure}{Figure}%
  \setautorefname{subfigure}{Figure}%
  \setautorefname{table}{Table}%
  \setautorefname{subtable}{Table}%
  \setautorefname{equation}{Equation}%
  \setautorefname{theorem}{Theorem}%
  \setautorefname{evaluation-question}{EQ\kern-\fontdimen2\font}%
  \setautorefname{lstlisting}{Listing}%
  \setautorefname{algorithm}{Algorithm}%
  \setautorefname{Item}{Item}%
  \setautorefname{item}{Item}%
  \setautorefname{footnote}{Footnote}%
  \setautorefname{page}{Page}%
}
\makeatother

\DeclareUnicodeCharacter{03B1}{\ensuremath{\alpha}}      %
\DeclareUnicodeCharacter{03B2}{\ensuremath{\beta}}       %
\DeclareUnicodeCharacter{2192}{\ensuremath{\shortrightarrow}} %
\DeclareUnicodeCharacter{2115}{\ensuremath{\mathbb{N}}}  %
\DeclareUnicodeCharacter{2264}{\ensuremath{\leq}}        %
\DeclareUnicodeCharacter{22A5}{\ensuremath{\bot}}        %

\newcommand{\ourlib}{Basalt\xspace}

\ifdraft

\newcommand{\todo}[1]{{\color{red}[TODO: #1]}}
\newcommand{\hg}[1]{\leavevmode{\color{purple}[HG: #1]}}
\newcommand{\mwh}[1]{{\color{blue}[Mike: #1]}}
\newcommand{\sem}[1]{{\color{magenta}[Segev: #1]}}
\newcommand{\kmh}[1]{{\color{orange}[Katie: #1]}}
\newcommand{\en}[1]{{\color{green!50!black}[Ernest: #1]}}   %
\newcommand{\jc}[1]{{\color{cyan}[Josh: #1]}}
\else

\newcommand{\todo}[1]{\ignorespaces}
\newcommand{\hg}[1]{\ignorespaces}
\newcommand{\mwh}[1]{\ignorespaces}
\newcommand{\sem}[1]{\ignorespaces}
\newcommand{\kmh}[1]{\ignorespaces}
\newcommand{\en}[1]{\ignorespaces}
\newcommand{\jc}[1]{\ignorespaces}
\fi

\newtheoremstyle{evaluationquestion}
  {\topsep}{\topsep}{\itshape}{}{\bfseries}{.}{.5em}
  {\thmname{#1}\thmnumber{#2}\thmnote{ (#3)}}
\theoremstyle{evaluationquestion}
\newtheorem{evaluation-question}{EQ}
\theoremstyle{acmplain}
\newtheorem*{theorem*}{Theorem}

\usepackage[most]{tcolorbox}
\newtcolorbox{evaluation}{%
  breakable, enhanced, blanker, parbox=false,
  left=0pt, right=0pt, top=2pt, bottom=2pt,
  before skip=\medskipamount, after skip=\medskipamount,
  borderline west={0.8pt}{-8pt}{black!35}}

\newcommand{\li}[1]{\leaninline{#1}}